\documentclass{acmart}
\setcopyright{rightsretained}

\copyrightyear{2026}
\acmYear{2026}
\acmConference[CHI '26]{CHI Conference on Human Factors in Computing
Systems}{April 13-17, 2026}{Barcelona, Spain}
\acmBooktitle{CHI Conference on Human Factors in Computing Systems (CHI '26), April 13-17, 2026, Barcelona, Spain}
\acmDOI{10.1145/XXXXXXX.XXXXXXX}
\acmISBN{XXX-X-XXXX-XXXX-X/XX/XX}

\usepackage{fontawesome}
\usepackage{algorithmic}
\usepackage{algorithm}
\usepackage{array}
\usepackage{url}
\usepackage{verbatim}
\usepackage{graphicx}
\usepackage{booktabs}
\usepackage{newfloat}
\usepackage{listings}
\newfloat{lstfloat}{tbp}{lop}
\floatname{lstfloat}{Listing}
\usepackage{support-caption}
\usepackage{multirow}
\usepackage{tabularx}
\usepackage{subcaption} 
\DeclareCaptionSubType{lstfloat}
\usepackage{mathtools}
\usepackage{microtype} 
\usepackage{tikz}
\usetikzlibrary {graphs, calc}
\usepackage{xspace}
\usepackage{comment}
\usepackage{siunitx}
\usepackage{enumitem}

\newcommand{\eg}{e.g.}
\newcommand{\ie}{i.e.}
\newcommand{\etal}{et~al.\xspace}
\newcommand{\IT}{Icicle tree\xspace}

\newcommand{\comma}{\,}

\newif\ifanon
\anonfalse

\ifanon
\newcommand{\anonurl}[1]{\texttt{anonymized.for.submission}}
\else
\newcommand{\anonurl}[1]{\url{#1}}
\fi

\iftrue  
\newcommand{\todo}[1]{\textcolor{red}{todo: #1}}
\newcommand{\jdf}[1]{\textcolor{black}{#1}}
\newenvironment{jdfenv}{\begingroup\color{black}}{\endgroup}
\newenvironment{aaenv}{\begingroup\color{blue}}{\endgroup}
\newcommand{\mi}[1]{\textcolor{blue}{#1}}
\newcommand{\am}[1]{\textcolor{black}{#1}}
\newcommand{\miC}[1]{\textbf{[-Mickaël-~}\mi{#1}\textbf{~]}}

\else
\newcommand{\todo}[1]{} 
\newcommand{\jdf}[1]{#1}
\newenvironment{jdfenv}{}{}
\newcommand{\mi}[1]{#1}
\newcommand{\am}[1]{#1}
\newcommand{\miC}[1]{}
\fi

\begin{document}

\title{ParcoursVis: Visualization of Electronic Health Record Sequences at Scale}



\author{Ambre Assor}
\email{ambre.assor@inria.fr}
\orcid{0000-0002-1825-0097}
\author{Mickael Sereno}
\email{mickael.sereno@inria.fr}
\orcid{000-0003-1298-0774}
\author{Jean-Daniel Fekete}
\email{jean-daniel.fekete@inria.fr}
\orcid{0000-0002-1825-0097}
\affiliation{%
  \institution{Univ.\ Paris-Saclay and Inria}
  \city{Saclay}
  \country{France}
}

\renewcommand{\shortauthors}{Assor et al.}

\begin{abstract} 
We present ParcoursVis, an open-source Progressive Visual Analytics tool designed to explore aggregated electronic health record sequences of patients at scale.
Existing tools are limited to about 20k patients that they can process fast enough to remain interactive, under human latency limits.
They need to process the whole dataset before showing the visualization, taking a time proportional to the data size.
Yet, managing large datasets allows for discovering rare medical conditions and unexpected patient pathways, contributing to improving treatments. 
To overcome this limitation, ParcoursVis relies on a \emph{progressive aggregation algorithm} that quickly computes an approximate initial result, visualized as an \IT, and improves it iteratively, until the whole computation is done. 
With its architecture, ParcoursVis remains interactive while visualizing the sequences of millions of patients---three orders of magnitude more than similar tools.
We describe our PVA architecture, which achieves scalability with fast convergence and visual stability.
\end{abstract}

\begin{teaserfigure}
\centering
\begin{tikzpicture}
  \node[anchor=south west,inner sep=0] (image) at (0,0) {
    \includegraphics[width=0.9\textwidth]{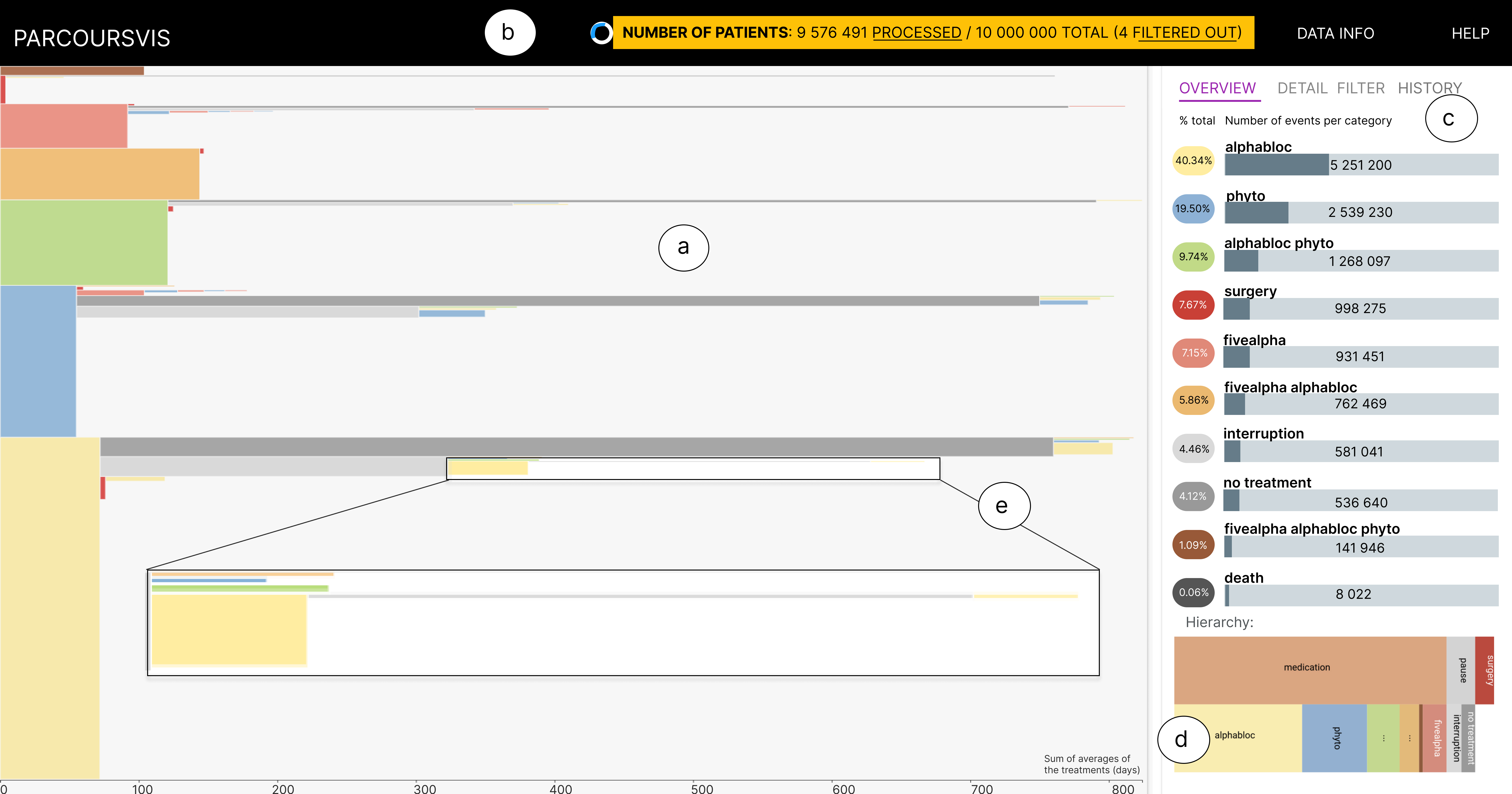}};
  \begin{scope}[x={(image.south east)},y={(image.north west)}]
  \end{scope}  
\end{tikzpicture}
  \caption{The ParcoursVis system. 
      (\subref{fig_mainView_a}) The main view shows 10\space{M} medical pathways aggregated, visualized as an \emph{\IT}, for patients treated for non-cancerous prostate adenoma.
      (\subref{fig_mainView_b}) The number of patients visible and filtered out.
      (\subref{fig_mainView_c}) The control panel with the legend and overall distribution of event types (e.g., buying medication or having surgery).
      (\subref{fig_mainView_d}) The hierarchy of event types, used to select a level of detail of the tree.
      (\subref{fig_mainView_e}) Zoomed view of a sub-tree, revealing more of the hierarchical structure.
      Each colored node represents a stage in a sequence of events starting from the root (left), flowing to the right.
      Child nodes representing the next stages are on the right of their parent, sorted by frequency (i.e., proportion of patients) from bottom to top.
      Node height is proportional to the frequency of the event, and width to the average duration (in days).
      Here, most patients started with the \emph{alphabloc} treatment (bottom left). 
      }
    \label{fig_mainView}
    \begin{subcaptiongroup}
        \phantomcaption\label{fig_mainView_a}
        \phantomcaption\label{fig_mainView_b}
        \phantomcaption\label{fig_mainView_c}
        \phantomcaption\label{fig_mainView_d}
        \phantomcaption\label{fig_mainView_e}
    \end{subcaptiongroup}
    \Description{Screen capture of the ParcoursVis system with five parts highlighted: (a), (b), (c), and (e).}
\end{teaserfigure}

\maketitle



\section{Introduction}

Visualization scalability is particularly challenging in healthcare, where Electronic Health Records (EHRs) can contain a large amount of multivariate temporal data for millions of patients over decades. 
As datasets increase in size and complexity, traditional visualization techniques struggle to maintain interactivity and responsiveness, thereby hindering effective exploratory data analysis on these large datasets. 
Indeed, results from research in visualization show that user attention declines when latency exceeds 500\comma ms, and after five to ten seconds, users tend to abandon tasks or lose focus~\cite{Zgraggen:2016_TVCG, Liu:2014_TVCG}.
Addressing the scalability challenge necessitates improvements in algorithms and the development of optimized data structures within visualization systems, but even with the heaviest optimizations~\cite{Mosaic, imMens}, existing visualization systems' latency remains proportional to dataset size, limiting their scalability.
To overcome this limitation, Progressive Visual Analytics (PVA) appears as a novel programming paradigm that decouples data size from latency~\cite{PDABook} by quickly computing and visualizing an approximate initial result and improving it iteratively until the whole computation is done.
Users can interact with the visualization at any step with low latency.
PVA has been effectively applied to several application domains~\cite{Ulmer:2024_TVCG}, but it introduces new challenges, such as a new programming architecture, as well as visualization stability, uncertainty management, and potentially usability issues, that should be considered for each new application, and in particular for the exploration of large EHR datasets.

For instance, the SNDS (the French national health database \cite{SNDS}) holds decades of health data on all the French citizens, encompassing records of drug purchases, medical diagnoses, and hospital treatments. Exploring such data has the potential to advance healthcare by enabling a better understanding of treatment practices and patient outcomes.
This approach aligns with the goal of medical professionals to assess whether patients' treatments adhere to national guidelines and, if deviations occur, to understand the reasons. This is key to providing patients with the highest quality of care through evidence-based medicine and verified recommendations.
Yet, the volume of such datasets necessitates innovative visualization approaches capable of handling large medical data while preserving efficiency and usability.

In this context, we developed ParcoursVis (\autoref{fig_mainView}), a PVA tool created in collaboration with medical institutions to visualize patient care pathways derived from EHRs.
Previous work has laid a foundation for similar visualization tools aimed at healthcare analysis (\eg,~\cite{Bjarnadottir:2016_PharmacoEconomics, Meyer:2013_VAHC, Ozkaynak:2015_BI}). For instance, EventFlow~\cite{Monroe:2013_TVCG} aggregates event sequences into a \emph{prefix tree} and visualizes it as an \emph{\IT}, as shown in~\autoref{fig_mainView}.
This visualization has proven effective for several applications (\eg,~\cite{Bjarnadottir:2016_PharmacoEconomics, Monroe:2013_TVCG, Ozkaynak:2015_BI}). 
We, therefore, build on the EventFlow visualization technique for ParcoursVis, along with several improvements. 

However, tools like EventFlow have generally been applied to relatively small datasets, sometimes encompassing fewer than 100 patients. According to our experiments, \jdf{EventFlow remains interactive up to about 20\comma k patients.}
While such tools have offered valuable initial insights, they fall short in scalability, which has been reported as a main challenge for EHR visualization~\cite{Wang:2022_CGF, 10.1093/jamia/ocae249}.
In practice, regional and national EHR databases present data at a far greater scale, often involving millions of patients. To our knowledge, no existing tool enables the visualization and interactive analysis of more than about 100\comma k EHR aggregated event sequences~\cite{Wongsuphasawat:2011_CHI, Wongsuphasawat:2012_TVCG, Monroe:2013_TVCG, Wang:2022_CGF}, which limits the scope of exploration. This constraint impedes the ability to gain a more comprehensive view of patient pathways, identify patterns and trends that may not be visible in smaller or sampled datasets, and discover rare patterns and outliers.

With ParcoursVis, we allow visualizing temporal event sequences of more than 100\comma{M} patients, potentially supporting the EHRs of the largest countries.
Our progressive approach provides a quick, approximate, yet \jdf{accurate} preview of the data early on and quick updates thereafter, converging within seconds. After the initial preview, our algorithm continues processing the dataset, updating the visualization every second or so until it is fully processed.

Yet, the progressive updates of the aggregated tree can cause visual instability, as the nodes (\ie, rectangles on~\autoref{fig_mainView_a}) are laid out at each update. The shifting of nodes can disrupt the viewer's mental map and create distracting visual flickers. Therefore, our approach may have an impact on the usability of the tool. To mitigate this issue, we introduce a \emph{sorting with hysteresis} algorithm that limits this flicker during progressive rendering.

In this work, we describe our PVA architecture and explain the changes to turn a traditional architecture for visualizing aggregated sequence data into a progressive one.

%

In summary, this article presents our system and reports on its efficiency in terms of PVA, stability, and possibly flicker. Specifically: 
\begin{enumerate}[nosep]
    \item We describe our PVA architecture to achieve scalability for aggregated event-sequence visualization over the web.
    \item We introduce our algorithm, specifically designed to process events progressively in order to provide meaningful intermediate results, and we evaluate its efficiency using metrics specific to PVA~\cite{Richer:2022_TVCG}. In particular, we report on \textbf{scalability}, and \textbf{latency}.
    \item To optimize visual stability despite the potential flicker introduced by PVA, we introduce a sorting algorithm with hysteresis, i.e., maintaining node order when size differences remain below a threshold. We evaluate \textbf{stability} through quantitative measures, showing that hysteresis-based sorting improves both stability and convergence of the progressive visualization.
    \item  We report on \textbf{usability} regarding the progressive aspect. 
    Our professional participants have completed a variety of tasks seamlessly, without mentioning any delays in the system's response, or discomfort with the progressive rendering of the \IT. 
\end{enumerate}
\am{
\textbf{We address a key challenge in aggregated EHR sequence visualization and present ParcoursVis wich improves scalability by at least three orders of magnitude compared to state-of-the-art systems.}
Central to this contribution, we present the first PVA tree visualization algorithm (according to the survey of Ulmer et al.~\cite{Ulmer:2024_TVCG}), and building on the results of PVA systems in other domains~\cite{Ulmer:2024_TVCG}, we show that ParcoursVis maintains interactivity, usability, and utility while ensuring scaling to much larger datasets. 
Ultimately, our goal is for ParcoursVis to provide a user experience comparable to EventFlow, but capable of handling up to 10,000 times more patients without users perceiving the progressive nature of the system.}The source code and evaluation scripts are available at \anonurl{https://gitlab.inria.fr/aviz/parcoursvis/}.



\section{Related Work}
ParcoursVis reuses the visualization technique of EventFlow because it was thoroughly tested and validated for analyzing medical temporal event sequences (\eg, ~\cite{Bjarnadottir:2016_PharmacoEconomics, Monroe:2013_TVCG, Ozkaynak:2015_BI}). Thus, our research goal is not to propose a new visualization technique but rather to explore the strategies to scale and adapt existing approaches that have already demonstrated their effectiveness and apply them to the largest EHR data available.
\am{Especially, we consider the usability of EventFlow's visualization validated and only evaluate ParcoursVis for the specific features it introduces, specifically the use of a progressive algorithm to ensure scalability and our strategy to avoid subsequent instability.}
Therefore, this work relates to temporal event sequences, scalability, PVA systems, and progressive uncertainty in the partial results.

\subsection{Temporal Event Sequences}
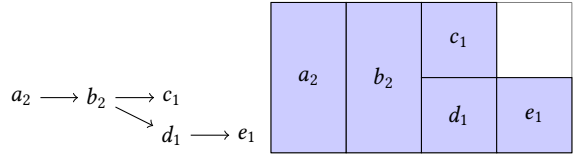
\begin{figure}[tb]
    \centering
    \begin{tikzpicture}
        \graph [branch down=5mm, math nodes] { a_2 -> b_2 -> { c_1, d_1 -> e_1} };
    \end{tikzpicture}
    \begin{tikzpicture}[fill=blue!20]
        \draw[help lines] (0, 0) grid (4,2);
        \path (0.5,1) node(a) [rectangle,minimum height=2cm,minimum width=1cm,draw,fill]    {$a_2$}
              (1.5,1) node(b) [rectangle,minimum height=2cm,minimum width=1cm,draw,fill]    {$b_2$}
              (2.5,1.5) node(c) [rectangle,minimum height=1cm,minimum width=1cm,draw,fill]  {$c_1$}
              (2.5,0.5) node(d) [rectangle,minimum height=1cm,minimum width=1cm,draw,fill]  {$d_1$}
              (3.5,0.5) node(e) [rectangle,minimum height=1cm,minimum width=1cm,draw,fill]  {$e_1$};
    \end{tikzpicture}
    \caption{ESS Aggregation of EHR Sequences as implemented by EventFlow and ParcoursVis. Patient \textit{A} took the treatment sequence $\{a, b, c\}$ in that order, and patient \textit{B} took $\{a, b, d, e\}$. The left side shows the \emph{prefix tree} of these sequences, and the right side shows the \IT visualization where node height encodes the frequency. 
    \am{Node width is constant here, but, in EventFlow and ParcoursVis, it encodes the mean duration of the treatments.}} 
    \label{fig_ESS}
    \Description{TODO}
\end{figure}

\begin{table}\ttfamily\small
\begin{tabular}{llllll}
\toprule
\textbf{id} & \textbf{age} & \textbf{type} & \textbf{date} & \textbf{drug} & \textbf{comorbidities}\\
\midrule
0 & 71 & fivealpha & 2015-05-29 & dutastéride & Diabetes \\
0 & 71 & fivealpha & 2015-06-05 & dutastéride & Diabetes  \\
1 & 50 & alphabloc & 2003-09-26 & alfuzosine & Hypertension \\
1 & 50 & alphabloc & 2003-10-03 & alfuzosine & Hypertension \\
1 & 50 & alphabloc & 2003-10-10 & alfuzosine & Hypertension \\
...\\
\bottomrule
\end{tabular}

\caption{CSV file showing records using our format, typical of event sequence systems. Patient \#0 is 71 and bought the drug ``Dutastéride'' containing the 5-alpha molecule in May and June 2015. He suffers from diabetes.
}\label{tab_csv}
\end{table}
Our tool, ParcoursVis, aims to support domain experts in visualizing and exploring patients' care pathways. 
Like EventFlow~\cite{Monroe:2013_TVCG}, it uses aggregation and filtering to visualize event sequences \am{such as the ones shown on \autoref{tab_csv}, extracted from the SNDS~\cite{SNDS}.}  
According to the taxonomy in Wang \etal's survey of EHR visualization~\cite{Wang:2022_CGF}, it falls into the ``Event Sequence Simplification'' (ESS) family of techniques (explained in \autoref{fig_ESS}).
This family is defined as ``any technique used for reducing the visual complexity of event sequences in aggregated display overviews''~\cite{Wang:2022_CGF}.
Wang \etal distinguish seven families: ML, NLP, ESS, Geospatial Visualization, Clustering, Comparison, and Others.
Among the 51 articles related to the visual analysis of EHR data, 11 belong to the ESS family.
In other families, visualizing an overview of the event sequences is not the focus; they rather focus on one patient, a set of patients, or spatial areas sharing some characteristics.

In addition to visualization, EHR data is also analyzed using \emph{sequence mining} augmented with visualization~\cite{DBLP_journals/tvcg/VrotsouN19, DBLP_journals/tvcg/LiuWDHWW17, DBLP_conf/iui/PererW14, Stolper:2014_TVCG}. The idea consists of trying to automatically discover frequent and important patterns in the sequences to discover unexpected information. This approach differs from ours since it only needs to visualize the search results. 
Instead, our tool interactively visualizes all the treatment pathways following Shneiderman's mantra: overview first, zoom, filter, and details on demand~\cite{Mantra}.

\subsubsection{Aggregation of EHR Data}\label{sec_aggregation}

\autoref{fig_ESS} explains how a list of sequence of events is aggregated into a tree. In addition, there is also an aggregation at the sequence level, applied by both EventFlow and ParcoursVis for creating \emph{high-level events}, which are the ones used to build the prefix tree.

An EHR database, like the SNDS~\cite{SNDS}, contains low-level events associated with each patient. 
For example, a patient treated for hypertension needs to buy a box of pills containing a specific molecule every three months. \am{Low-level events in the SNDS database~\cite{SNDS} include, for instance, the drug's date of purchase and the associated molecule (~\autoref{tab_csv}).  
When the patient purchases the same box repeatedly over time (it can be up to many years), as shown for patients 0 and 1 on \autoref{tab_csv}, we transform the sequence into one high-level event (refered simply as an \emph{event}) with a start-date and end-date (\ie, a duration) associated with the active molecule, whatever the brand. Here, patient 0 experiences the \emph{event} ``fivealpha'', with a duration of 6 days (from 2015-05-29 to 2015-06-05).}
This kind of aggregation rule simplifies massively the prefix tree and is standard in EventFlow and ParcoursVis.

ParcoursVis can handle more complex aggregation rules for mixing drugs or generating synthetic events when treatments are interrupted or abandoned. 
For example, in the case of prostate adenoma treatment, shown on \autoref{fig_mainView_c}, patients can take different types of molecules at the same time, such as alphabloc and fivealpha. The visualization builds and shows the event ``fivealpha alphabloc'' accordingly.
In other contexts, two consecutive low-level events with different molecules are interpreted as switching from one treatment to another.
We also create \emph{synthetic events} to track \emph{adherence to treatment} when a patient takes too long before buying the next box of treatment (``interruption'' event), or stops their medication (``no more treatment'' event). When setting up the visualization for a particular pathology or use case, an expert needs to extract the relevant low-level events and specify the different rules to apply. 

\am{Given the large number of events available in databases such as the SNDS~\cite{SNDS}, which hosts the medical records of the entire French population, 
\jdf{designing their visualization to remain interactive at scale is a challenge}. Next, we introduce what scalability is and how to characterize it, and review the state of scalability in related applications.}

\subsection{Scalability}

\subsubsection{Progressive Visual Analytics}\label{sec_PVA}

PVA is the key technique we use to guarantee that ParcoursVis is interactive for very large dataset sizes.
PVA allows users to visualize the result of an algorithm while it is computed, in a progressive way. Instead of processing the whole data at once, waiting an unpredictable time, as standard algorithms do, PVA algorithms process data chunk by chunk (or iteration by iteration in some cases)~\cite{PDABook} to control the latency and show intermediate results iteratively at a controlled pace. 
The main goal is to keep the overall system interactive and yield results within the typical user's attention span that lasts 1--10 seconds~\cite{Nielsen:1993_Book, Zgraggen:2016_TVCG}. Practitioners already applied and studied the PVA paradigm on multiple types of visualizations, as described in the survey of Ulmer \etal~\cite{Ulmer:2024_TVCG}. 

In this work, we progressively aggregate event sequences of patients' care pathways into a prefix tree~\cite{Wang:2022_CGF} that we visualize as an \IT. 
Although PVA has been used for pattern mining in EHRs~\cite{Stolper:2014_TVCG}, to the best of our knowledge, we are the first to apply PVA to visualize EHR aggregated data at scale.
We also introduce the first progressive tree visualization technique, according to the survey of Ulmer \etal

By quickly presenting partial results, PVA systems allow users to get quick feedback on the overview and queries; \emph{it separates latency from completion time}.
It was also shown to increase users' commitment and actions per minute~\cite{Zgraggen:2016_TVCG}. 
In the next subsection, we describe the scalability characterization introduced by Richer \etal~\cite{Richer:2022_TVCG}. Based on their definition, we show, in this article, that our PVA algorithm for computing the prefix tree is indeed scalable: the aggregation time scales linearly with the number of events, inversely proportional to the number of cores, and updates always meet the requested latency. 

\subsubsection{Modeling Scalability}\label{sec_scalability}
Richer \etal~\cite{Richer:2022_TVCG} characterize scalability in visualization using a conceptual model that expresses the scalability of a visualization system by a function $f:(S; R, A) \longmapsto E$ with four components: the \emph{problem size $S$} (i.e., set of variables attuned to the size of the input data), \emph{resources $R$} (i.e., hardware or application environment used for the visualization), \emph{assumptions $A$} (\ie, boundaries of the research context and problem definition), and \emph{effort $E$} (i.e., measures of the claimed system's performance).

In our case, $S$ relates to the number of patients and the number of events per patient.
The two numbers are proportional if we consider patients to have an average number of events, but depending on the pathology studied, the average can vary from ten (\eg, for emergency service sequences) to thousands (\eg, for lifelong chronic pathologies).
$R$ is related to the machine we use and the number of CPU cores used by the computation.
Assumptions $A$ appertain to (1) the data, (2) the technical setup, and (3) the preprocessing.

\begin{enumerate}[nosep]
    \item We limited the number of event-types (e.g., ten in \autoref{fig_controlpanel}, between eight and twenty in other applications) since event types are encoded with color, and maintaining a relatively small set ensures that the primary events remain sufficiently distinguishable.
Note that the event type distribution is always skewed, as shown in \autoref{fig_controlpanel_c}.
    \item  We use a web-based application where a back-end machine serves the aggregated data through the internet, featuring a network bandwidth that can be low due to hospital wifi congestion, and a front-end local browser to interact with the visualization. We exclude the use of a GPU in the browser since we do not control the front end machines.
    \item  We also perform some preprocessing ahead of time in the back-end, without precomputing all the possible interactive queries, as our datasets can be filtered on a large combination of attributes.
As for $E$, we consider three efforts: \emph{latency}, \emph{completion time}, and \emph{stability}. 
Note that in non-progressive applications, latency is equivalent to completion time and that stability is specific to PVA and streaming visualizations~\cite{Ulmer:2024_TVCG} (see \autoref{sec_PVA} and \autoref{sec_uncertainty}).
\end{enumerate}


The scalability model of Richer \etal\ also introduces multiple ways of characterizing the scalability they call \emph{expression and meaning}, such as asymptotic scalability using the $O(\cdot)$ notation, or simply claiming to be faster than another system.
In our case, \emph{Latency} should always be under a specified limit to guarantee that users do not wait during the progressive updates of the system, and completion time should be minimized. Also, during the progressive updates, we want to maximize the stability of the visualization to allow users to scan the \IT without sudden changes. 
We measure the first two efforts by varying the size $S$ of a test dataset and the number of CPU cores $C$, obtaining a \emph{scalability shape}, \ie, a graph $T=f(S, C)$.
The latency between two updates should remain under the 2\comma s boundary for all problem sizes and resources ($S$ and $C$).
Since the algorithm needs to process all the events, the completion time cannot be better than linear with $S$ and $C$, and short.
As for the stability, it can be measured in terms of variations or accidents per unit of time, but also as feedback from users because, ultimately, the stability might be considered good enough if the users do not notice instability. We report on both measures.




\subsubsection{Scalability in EHR Visualization}

\newcommand{\NA}{{Unspecified}}
\begin{table}
    \centering
    \begin{tabular}{l r l l}
    \toprule
    Study & Reported size & Unit & Scalability claims\\
    \midrule
    Bernard \etal~\cite{bernard2015visual} & 65 & patients & \NA \\
    Klemm \etal~\cite{klemm20153d} & 1,186 & subjects & \NA \\
    Perer \etal~\cite{perer2014frequence} & 2,336 & patients & \NA \\
    Gotz \etal~\cite{gotz2014decisionflow} & $2,899$ & patients & \NA \\
    Guo \etal~\cite{GuoEventThread} & 5,800 & patients & \NA \\
    Monroe \etal~\cite{Monroe:2013_TVCG} & $\approx 10,000$ & patients & Latency (same as Time) \\
    Jiang \etal~\cite{jiang2016healthcare} & 833,710 & cases & \NA \\
    \textbf{ParcoursVis} & \textbf{$> 20,000,000$} & \textbf{patients} & \textbf{Latency, Time, Stability} \\
    \midrule
    Vrotsou \& Nordman~\cite{DBLP_journals/tvcg/VrotsouN19} & 6,477 & sequences & \NA \\
     & 989,925 & sequences  & \NA \\
    Liu \etal~\cite{liu2016patterns} & $\approx 100,000$ & sequences  & \NA \\
    \bottomrule
    \end{tabular}
    \caption{Reported dataset sizes and scalability claims for the visualization of EHR and other sequence data.}
    \label{tab_reported}
\end{table}
Existing systems that visualize event sequences, both in medical and non-medical domains, often fail to report their scalability relative to the dataset sizes they handle~\cite{Richer:2022_TVCG}.  Wang \etal's survey~\cite{Wang:2022_CGF} does not report dataset sizes, although it reports scalability as the first future research challenge.
In the medical domain, reported dataset sizes vary widely, as shown in \autoref{tab_reported}.
EventFlow handles datasets with a limited number of sequences~\cite{Bjarnadottir:2016_PharmacoEconomics, Monroe:2013_TVCG, Ozkaynak:2015_BI, Du:2017_TVCG}; about 20\comma k and with a few seconds latency according to our experiments (available in the \hyperref[sec_compareEventFlow]{Supplemental Material}).
For handling larger datasets, its manual suggests the strategy of random sampling sequences~\cite{Du:2017_TVCG}. 
Random sampling, however, loses rare and sometimes important sequences that interest practitioners. 
With ParcoursVis, we found sequences concerning 3\comma k patients in our multi-million patient datasets; they are relatively rare, but not considered outliers by doctors.
Instead, to be scalable and accurate, we use PVA.
Some other systems, such as DecisionFlow~\cite{gotz2014decisionflow}, mention handling over a million individual point events (2,899 patients). However, they provide no details on their scalability.
In non-medical domains (mostly website visit sequences), studies report dataset sizes in the range of thousands to millions of sequences, but provide no scalability measure. 
Overall, for systems dealing with event sequence data, there is either a lack of information on performance relative to dataset size or a focus on small datasets, making it difficult to assess scalability \am{and leaving open the question of how such systems can handle large-scale, real-world event sequence data.}
\am{To exemplify the scalability of our approach}, we visualized the largest dataset mentioned with ParcoursVis (see the \hyperref{other_domains}{Supplemental Material}). 

\am{In contrast to event sequences, tabular data visualization systems provide well-documented scalability benchmarks. They typically rely on online analytical processing (OLAP) over one or several dimensions using aggregation operators such as mean, variance, sum, or count. Some systems, such as imMens~\cite{imMens} (which relies on GPU acceleration) and Mosaic~\cite{Mosaic} (which relies on the optimized analytical database DuckDB~\cite{DuckDB}), scale to billions of rows. However, event sequence aggregation relies on \jdf{a different data model and} more complex aggregation rules, as explained in \autoref{sec_aggregation}, rather than these generic operators, making OLAP techniques incompatible with our data. Consequently, their scalability claims are not directly comparable to those for event sequence visualization systems.}

\subsection{Uncertainty and Stability}\label{sec_uncertainty}

PVA systems start by showing partial results, implying uncertainty due to the progressive aspect of the computation. This \emph{progressive uncertainty}~\cite{PDABook:6} is reduced over time until the final result is computed, without any progressive uncertainty but potentially with other types of uncertainty, just like standard data and visualizations.
The visualization shown early may or may not be close to the final result. It is, therefore, important for analysts to assess the quality of the progressive results to decide if they can trust the visualization or if they should wait longer.
Angelini \etal~\cite{Angelini:2018_Informatics} explain that progressive visualizations undergo three stages: (1) early partial results that are usually noisy and not trustworthy, (2) mature partial results that are stabilizing and can be trusted but with uncertainty, and (3) definitive partial results that are accurate and can take (wasted) time to finish.  When our aggregation tree is computed progressively, all the nodes will undergo these three stages. Yet, the nodes closest to the root will become reliable and stable earlier than the deeper nodes since they are aggregating more sequences, and that stability is essentially related to the number of sequences they represent. 

The uncertainty can be computed and visualized explicitly (\eg, through error bars), or implicitly through measures such as stability. Instability at a given stage indicates that the result is still uncertain, though the reverse is not always true: a progressive visualization may remain stable for some time and still change later~\cite{Patil:2023_TVCG, Correll:2014_TVCG}.
However, explicitely conveying uncertainty remains challenging, even for statisticians who often struggle to interpret it correclty in static visualizations~\cite{Correll:2014_TVCG}. For progressive visualization, many representations are possible, \eg,  bar charts with error bars, violin plots, or gradient distribution~\cite{Correll:2014_TVCG, Fisher:2012_CGA}, but few have been studied and validated empirically. Patil \etal~\cite{Patil:2023_TVCG}, for instance, evaluated four designs for visualizing uncertainty on progressive bar charts and found two to be effective. However, there is no study about the progressive visualization of an \IT, to the best of our knowledge.

ParcoursVis takes less than 20\comma s to update its tree fully for our 10\comma M patients dataset, with an update every 2\comma s.
It seems unlikely that users could make sense of any uncertainty visualization added to many nodes in such a short update time; it might take more time to make sense of the uncertainty visualization than to wait for the progressive visualization to complete.
Instead, we rely on stability as a proxy for quality, assuming that subtrees will remain stable when the progressive uncertainty decreases~\cite{PDABook:6}.
On the other side, when a subtree is not stabilized for a few seconds, the user will not be able to infer much from watching it. 
We study the convergence and stability of ParcoursVis's progressive \IT visualization, instability reflecting progressive uncertainty.

\section{ParcoursVis}
We first describe ParcoursVis to set the overall context of our work before digging into its PVA-related algorithms. 
We also share knowledge about the PVA's impact on the software structure that we discovered when building ParcoursVis.

ParcoursVis is a web-based application inspired by EventFlow that processes data in a back end; users interact via a web browser front end.
While ParcoursVis is specialized for exploring and analyzing EHR sequences, its core structure is more general and can handle the same kinds of event sequences and applications as EventFlow.
Contrary to EventFlow, the rules for aggregating and merging events are written in C++ for performance and expressive power reasons.
For real pathologies, many constants and aggregation rules need to be set according to state-of-the-art recommendations that cannot be expressed within EventFlow.
Therefore, an engineer is needed to configure ParcoursVis to define new domain-specific rules when it is applied to a new medical question with specific kinds of events.



\subsection{Example Use Case}
\am{ParcoursVis is currently used in two main applications: (1) non-cancerous prostate adenoma treatment analysis for the French social security, and (2) analysis of patient pathways across all public Parisian hospitals' Emergency Departments (EDs). While the detailed examples in this paper come from the first use case, we briefly describe the second.}

(1) Specific to our non-cancerous prostate adenoma use case, our application receives information from the Social Security reimbursement database (available in the SNDS~\cite{SNDS}) about the drugs bought by patients (time, name, molecule), dedicated surgery, and deaths.
As explained in \autoref{sec_aggregation}, there are three molecules possibly administered for treatment: alpha-blockers, specific herbal medicine, and 5-alpha, respectively labeled alphabloc, phyto, and fivealpha in~\autoref{fig_mainView_c}. 
Those are the three low-level events our referring doctors selected to extract for their analysis.  They also extract prostate surgery events and deaths. These events are aggregated into high-level events with durations that are transformed into an \IT and visualized. 
Our dataset also contains \am{patient's attributes such as} their age and information related to known \am{comorbidities} (\eg, diabetes, hypertension) to explore possible correlations with treatment pathways.

(2) We also use ParcourVis for visualizing patient pathways in the 16 public Emergency Departments in Paris. For over 15 years, EDs have faced a steady increase in patient volume, leading to frequent overcrowding and prolonged waiting times. Analyzing patient care pathways offers the potential to identify specific stages where the process becomes delayed, resulting in extended waiting times or medical accidents. Our project deals with 1.5\space{M} patient visits over three years. 
\am{To use the secure JupyterLab environment provided by the Paris Public Hospitals, we wrapped the ParcoursVis front end into a widget to provide an easy-to-access interface for our collaborators.}
\am{Emergency Department staff use a centralized system to document patients’ medical records upon their arrival. Physicians enter administrative information (\eg, name, gender, age) and medical procedures (\eg, triage score assessment, medical exams). They also document patient location within the hospital as patients are redirected to different stages of care (\eg, to a standard cubicle or waiting areas). This system saves data every 15 minutes, providing near-real-time location data on each patient throughout their stay. As in the prostate adenoma use case, low-level events (here, changes in patient location within the ED, recorded every 15 minutes) are aggregated into higher-level stages of care, such as, for instance, being located in a waiting area (where patients remain before consultations or results), the triage room (where priority scores are assigned) and the shock room (Intensive Care Unit).} \am{In addition to these locations, we include attributes such as patients' age, the name of the hospital they visited, and the triage score they have been given. The triage score, ranging from 1 (high priority, life-threatening) to 5 (low priority, non-urgent), is assigned to patients based on symptoms, medical history, vital signs, and clinical observations. 
}

%

\newcolumntype{Y}{>{\raggedright\arraybackslash}X} 
\setlength{\tabcolsep}{4pt} 
\renewcommand{\arraystretch}{1.1} 
\newcommand{\CHECK}{\faCheckSquareO}
\newcommand{\UNCHECK}{\faSquareO}

\begin{table}\small
\begin{tabularx}{\textwidth}{p{3.0cm}Ycc}
\toprule
\multicolumn{1}{l}{} & \textbf{Tasks} & \textbf{ParcoursVis} & \textbf{EventFlow} \\ \midrule

\multirow{4}{*}{\textbf{Simple Tasks}} &
\textbf{Overview:} visualize all the temporal event sequences (high-level event types). & \CHECK & \CHECK \\ \cmidrule(r){2-4}
& \textbf{Browse explicit sequences:} enable record-level inspection and drill down into individual records. & \UNCHECK & \CHECK \\ \cmidrule(r){2-4}
& \textbf{Zoom and Filter Events:} visualize all the temporal sequences having a given prefix (children of a selected node) or filter out specific event types. & \CHECK & \CHECK \\ \cmidrule(r){2-4}
& \textbf{Details on demand:} visualize all the attributes and distributions related to a specific sequence (e.g., age distribution and duration distribution of the patients reaching a node). & \CHECK & \CHECK \\ \midrule

\multirow{5}{*}{\textbf{Advanced Tasks}} &
\textbf{System Configuration:} the ability to configure the system to perform different aggregations or data preparation. & \UNCHECK & \CHECK \\ \cmidrule(r){2-4}
& \textbf{Filter by attribute:} simplify the visualization by filtering the data based on the attributes and metadata of the sequence (e.g., total duration, age of the patient), and the resulting tree (e.g., minimum size of the nodes). & \CHECK & \CHECK \\ \cmidrule(r){2-4}
& \textbf{Configure view:} setting an alignment prefix to visualize what had happened before and after a specific sequence. & \CHECK & \CHECK \\ \cmidrule(r){2-4}
& \textbf{Abstracting:} regrouping multiple types of events into a super-type, using a hierarchy. & \CHECK & \CHECK \\ \cmidrule(r){2-4}
& \textbf{History and comparison:} navigate through the history, i.e., previous filters, zooms, or alignments. Compare different trees corresponding to different history states by switching between them. & \CHECK & \UNCHECK \\ \cmidrule(r){2-4}
& \textbf{Temporal queries:} provide a temporal query language to search for \eg, the absence of treatment or nested treatments. & \UNCHECK & \CHECK \\
\midrule

\multirow{2}{*}{\textbf{Analytical Tasks}} &
\textbf{Application questions:} identify trends, discover frequent, critical, or surprising pathways, and understand underlying causes, finding actionable strategies for improvement or resolution. & \CHECK & \CHECK \\ \cmidrule(r){2-4}
& \textbf{Data questions:} identify data errors, limitations, and determine more suitable data extraction methods. & \CHECK & \CHECK \\ \bottomrule
\end{tabularx}
\caption{A summary of the different tasks handled by ParcoursVis, outlining the differences with EventFlow.
}\label{parcoursvis_eventflow}
\end{table}

\subsection{User Interface and Tasks}\label{sec_tasks}
\begin{figure}\centering
    \includegraphics[width=0.7\linewidth]{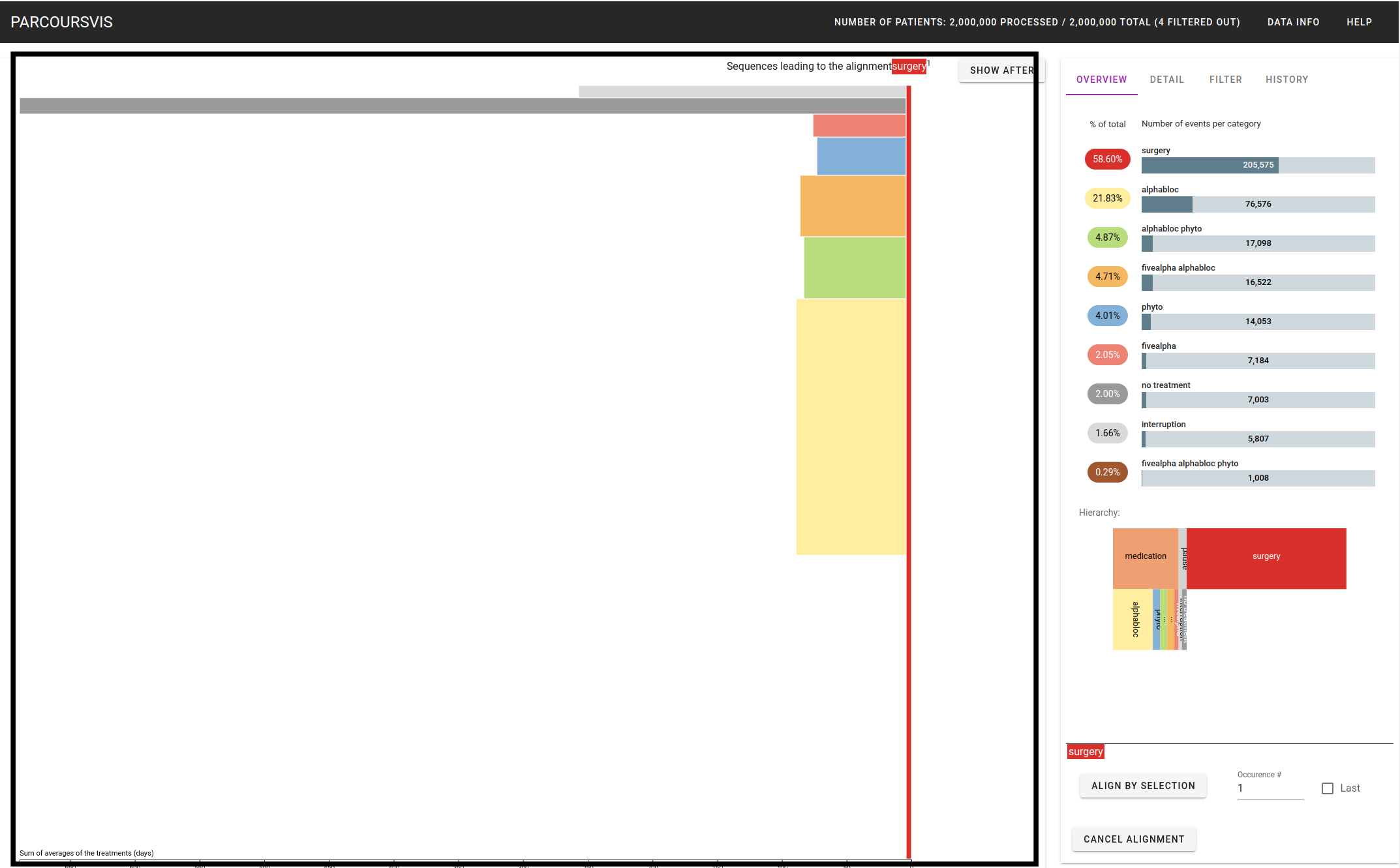} 
    \caption{The ``Sequence'' view visualization (squared) shows all events leading to a specified \emph{prefix}, here the ``surgery'' event. About 60\% of the patients who underwent surgery were treated before. A button allows showing all events that came after the prefix instead. }\label{fig_sequenceView}
    \Description{TODO}
\end{figure}
Our visualizations and user interfaces (GUI) are close to EventFlow, which has already been validated through many applications and case studies~\cite{Bjarnadottir:2016_PharmacoEconomics, Monroe:2013_TVCG, Ozkaynak:2015_BI}. 
\am{We summarize the different tasks supported in \autoref{parcoursvis_eventflow}.}
In the remaining, we focus on the specific features introduced by PVA\@.

\subsubsection{Main View and Prefix View}
The ``Main'' view (\autoref{fig_mainView_a}) shows the aggregated tree of all patients' care pathways. 
\am{The node width, height, and sorting order are explained in the Figure caption.}
The color of a node encodes its high-level event type. Clicking on a node shows details on demand in the ``Detail'' tab of the control panel (\autoref{fig_controlpanel_a}).

Users can change the view and focus on a specific event or sequence of events. The ``Prefix'' view (\autoref{fig_sequenceView}) shows the sequence tree that leads to or follows a given prefix, specified interactively.
Each of these views is updated progressively for scalability to guarantee interactive latency. 


\subsubsection{Control Panel}\label{sec_controlpanel} 
\begin{figure}
    \begin{subfigure}[b]{0.4\linewidth}
        \includegraphics[width=\linewidth]{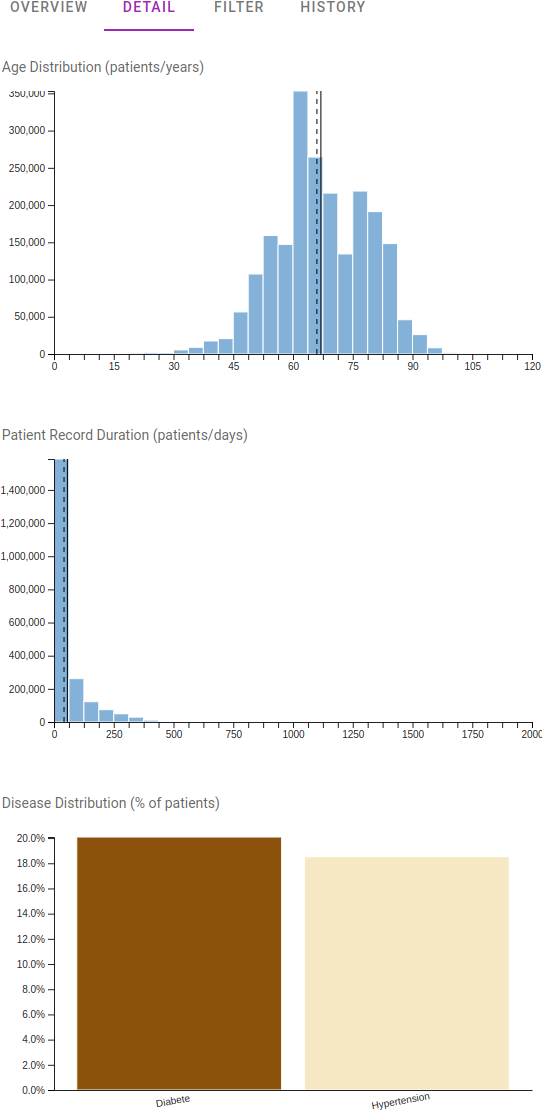}
        \caption{``Detail'' Tab showing histograms and bar charts for the selected node.}\label{fig_controlpanel_a}
    \end{subfigure}%
    \hspace{3mm}%
    \begin{subcaptiongroup}
        \begin{minipage}[b]{.4\linewidth}
            \centering
            \begin{minipage}[b]{\linewidth}
                \centering
                \includegraphics[width=\linewidth]{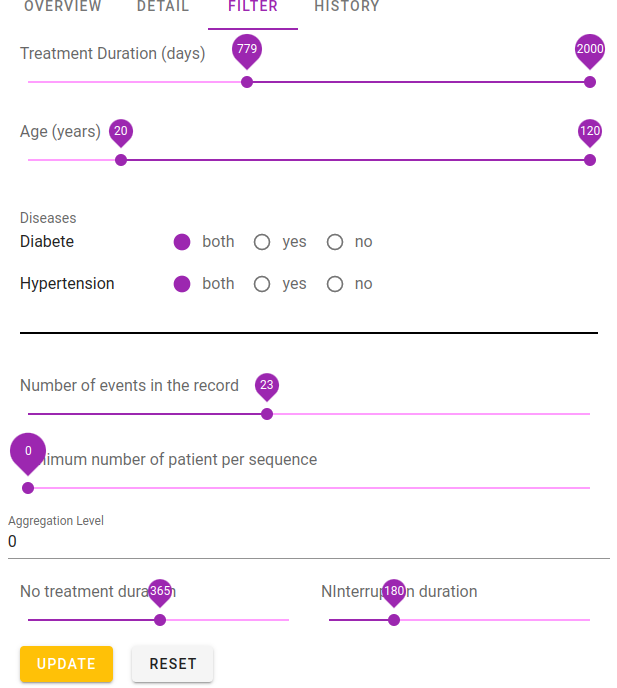}
                \caption{``Filtering'' Tab showing widget to set parameters and filter on attributes.}\label{fig_controlpanel_b}
            \end{minipage}
            \begin{minipage}[b]{\textwidth}
                \centering
                \includegraphics[width=\linewidth]{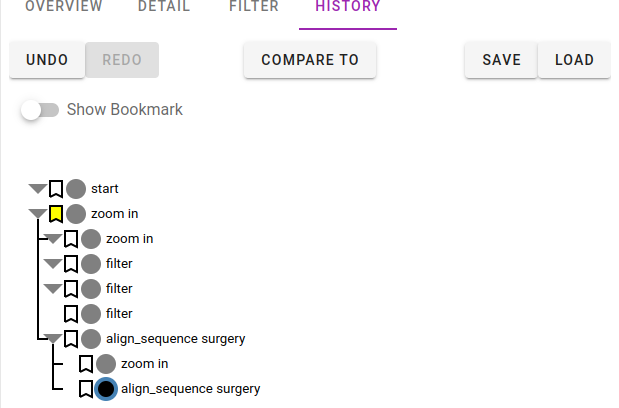}
                \caption{``History'' Tab showing the history tree for navigation and comparison. }\label{fig_controlpanel_c}
            \end{minipage}
        \end{minipage}
    \end{subcaptiongroup}
    \caption{Control Panel Tabs of ParcoursVis. See \autoref{fig_mainView_c} for the Overview Tab.}\label{fig_controlpanel} 
    \Description{Control Panel Tabs of ParcoursVis. (a) shows a histogram for age distributions, a histogram of durations, and a bar chart of comorbidities. 
    (b) show a set of control widgets for filtering over multiple attributes. (c) shows the navigation history tree to navigate in history and compare two points.}
\end{figure}
Besides the ``Main'' or ``Prefix'' views, the control panel shows details on demand and a dialog box to set a prefix on which to align the view. 

\begin{description}[nosep,leftmargin=0.5cm]
\item[\textbf{Overview}]
The ``Overview'' tab (\autoref{fig_mainView_c}) shows (1) the number of high-level events aggregated and (2) the hierarchy associated with the data using an \IT. Hovering a node shows the number of patients it contains. Those numbers are updated continuously when the progressive algorithm is running.
\jdf{Clicking on an event hides it from the tree. Clicking a second time makes it appear again. These operations are done in the front end and are instantaneous.}

The hierarchy groups multiple high-level events into a higher category, \eg, all drugs under the category ``medication''. \autoref{fig_mainView_d} visualizes the hierarchy and its distribution in the partial aggregated tree. For applications with many event types, using higher hierarchy levels simplifies the visualization. 
\am{Users can modify the level of hierarchy at any time; the change is instantaneous, as it only impacts the prefix tree visualization;}
\jdf{it does not need to restart the progressive aggregation.}

\item[\textbf{Detail}]
The ``Detail'' tab (\autoref{fig_controlpanel_a}) shows the data distributions of a selected node. Currently, we show the age and duration distributions using histograms and the comorbidities using a bar chart. Those graphs are updated iteratively when the progressive algorithm runs.

\item[\textbf{Filtering}]
The ``Filtering'' tab (\autoref{fig_controlpanel_b}) allows selecting a subset of patients and filtering the view according to parameters and node attributes. When users adjust the aggregating parameters (\eg, the thresholds defining the interruption events) or select a subset of patients by specifying query parameters, the progressive algorithm restarts from the beginning, considering only sequences matching the query and updating the rules that aggregate low-level events into high-level events. 
In contrast, filtering the view only manipulates nodes from the already computed tree (\eg, hide small nodes or merge nodes into higher hierarchical levels); it does not process the data again and results in instantaneous updates.
To filter the view, users can specify a hierarchical level, truncate the tree to a maximum depth (\eg, show only the first three depth levels), and select a minimum number of patients per sequence to hide small nodes, simplifying the tree.


\item[\textbf{History and Comparison}]
The ``History'' tab (\autoref{fig_controlpanel_c}) is inspired by the history comparison feature of Pister \etal~\cite{Pister:2023_CGF} that relies on the Trrack~\cite{Cutler:2020_VIS} history management library. It shows the history of the user's actions as a tree; the user can jump to any past state of the exploration by selecting a history node. 

The user can compare the current state with another one using the ``compare to'' button. 
The user can visualize the aggregated trees from both \jdf{states} in a back-and-forth manner using the ``Next'' and ``Previous'' buttons (not shown on the screenshots).
The node associated with the current state in the ``History'' tab has a blue halo, while the node compared has a green halo. 
\jdf{Selecting this second state initially restarts the progressive aggregation using the state's associated parameters.}
As the state being compared may not have finished processing all patients when switching back and forth between the compared states, \jdf{parcoursVis initializes a second progressive context for the second state. Thus, it does not restart the aggregation from the beginning, but resumes it from where it was left previously.} This avoids unnecessary waiting for the user to get back to where the context was previously and avoids breaking the user's mental model related to the context because the progressive algorithm may never get back to the exact previous view.
\end{description}

Overall, the PVA aspect of ParcoursVis  required us to consider the usability issues of every feature that either
(1) forces the aggregation algorithm to restart,
(2) conversely, avoids restarting the algorithm  when the front end can handle it alone, and 
(3) affects the user experience in an unexpected way (\eg, maintaining internally two prefix trees updated progressively for the comparison feature in the ''History`` tab).

\subsection{Implementation}

\begin{figure*}[tb]
    \centering
    \begin{tikzpicture}
        \tikzstyle{every node}=[font=\small]
        \node[anchor=west] (CSV) at (0,0) {\rotatebox{90}{CSV file}};
        \draw[->,dashed] (CSV.east) --+ (0.3, 0) node[rectangle,right,draw] (preparse) {Pre-processing};
        \draw[->,dashed] (preparse.east) --+ (0.3,0) node[rectangle,right,draw,solid] (memmap) {Binary File};
        \node[draw,dashed,rectangle,minimum width=4.0cm, minimum height=1.1cm] at ([xshift=0.5cm]preparse) {};
        \draw ([xshift=0.5cm]preparse |- 52,0.75) node [above,yshift=-0.2cm,align=center] {Pre-processing of the dataset.~\\Performed only once.};
        \draw[->] (memmap.east) --+ (1.4,0) node[midway,align=center] {Memory\\ Map} node[rectangle,right,draw,align=center] (backend) {PVA backend};
        \draw[<-,dotted] (backend.north) --+ (0,0.3) node[above,draw,rectangle,solid,align=center] {Hard-coded\\aggregating rules};


        \node[anchor=east,rectangle,draw] (UI)   at (14.5,0.35) {UI};
        \draw (UI |- 52,-0.35) node [rectangle, draw] (view) {View};
        \node[rectangle,draw] (frontend) at (13,0) {Front end};
        \draw[->] ([yshift=-2]frontend.east) -- (view.west);
        \draw[<-] ([yshift=2] frontend.east)  -- (UI.west);
        \draw[thick,<->] (backend.east) -- (frontend.west) node [midway,fill=white] (PVA) {\bfseries PVA pipeline};

      \path[every node/.style={font=\sffamily\small}]
          (frontend) edge[bend right=15,red,->] node [below,midway] {Filter Tree} (backend)
          (frontend) edge[bend right=20,->] node [above,midway] {Filter Data} (backend);

      \path[every node/.style={font=\sffamily\small}]
          (backend) edge[bend right=20,red,->] node [below,midway] {Update Tree} (frontend) 
          (backend) edge[bend right=15,<->] node [above,midway] {Details on Demand} (frontend);
    \end{tikzpicture}
    \caption{Architecture of ParcoursVis, organized in a pre-processing stage, a back end, and a front end. \jdf{The progressive management of the tree is depicted in red.}}
    \label{fig_parcoursvisArchitecture}
    \Description{TODO}
\end{figure*}
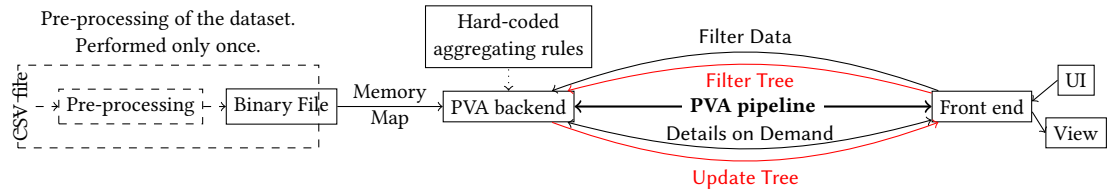

\autoref{fig_parcoursvisArchitecture} describes the architecture of ParcoursVis.
Our front-end application is made of about 3,000 lines of JavaScript that uses \texttt{D3.js}~\cite{D3} and \texttt{Vue.js}~\cite{Vue}. This allows us to show visualizations on a regular web browser accessible to anyone and, in particular, to referring doctors without any installation. However, web technologies are not designed for heavy computation. 
We, therefore, rely on an optimized back-end library written in about 3,000 lines of C++ and used as a Python module that progressively aggregates the data sent to the front end. The back-end Python server code is 127 lines long. We also provide test suites for JavaScript and C++ that double the source code size. The code is open-source at \anonurl{https://gitlab.inria.fr/aviz/parcoursvis/}.

While our main view shows aggregations and visualizations similar to EventFlow, our engineering design differs in many ways from it for scalability and stability purposes.

\subsection{Optimization of ESS Visualization Programs}



All ESS systems we know are implemented using a structure similar to the simplified code in~\autoref{lst_simple}. 
They take a file as input, usually in CSV format as shown in~\autoref{tab_csv} and produce a prefix tree that is then rendered on screen. 
Their run time is essentially linear in the number of events, but existing implementations use data structures that are expensive both in memory and execution time, limiting the dataset size to achieve the computation under the 1--10\comma s latency barrier. 
Improving the scalability can be done with three strategies: 
(1) optimizing the data structures and algorithms \faBarChart,
(2) parallelizing the program~$\rightrightarrows$, and 
(3) turning it into a progressive program \faSpinner. 
Only applying strategies 1--2, the function ``aggregate'' (line 15 of \autoref{lst_simple}) will continue to take a time proportional to the number of events. 
Above some number, the time will exceed the acceptable latency and degrade the interaction. 


ParcoursVis applies the three strategies, as shown in \autoref{lst_aggregation}. It optimizes the loading time, the data structures~\faBarChart, parallelizes the aggregation $\rightrightarrows$, and performs the computation in a progressive manner to decouple processing time from \jdf{rendering} latency~\faSpinner. 

\begin{lstfloat}
\begin{lstlisting}[language={Python},basicstyle=\ttfamily\small,escapechar=\%,morekeywords={\@dataclass,field}]
def ESS(filename):
    events = pd.read_csv(filename, index_col=None)
    events.sort_values(by=["id", "date"], inplace=True)
    patients = events["id"].unique()
    tree = aggregate(patients, events)
    render(tree)

%\label{lst_simple_node}%@dataclass class Node:
    type: str
    count: int = 0
    children: Dict[str, Node] = {}
    agedist: Counter = Counter()
    durationdist: List[int] = []

def aggregate(patients, events):
    tree = Node(type="root")
    for patient in patients:
        patient_events = events[events["id"]==patient]
        aggregate_patient(patient_events, tree)
    return tree

def aggregate_patient(lowlevel, tree):
    current_node = tree
    context = Context()
    for event in lowlevel+ENDEVENT:
        for hl in context.ll_to_hl(event):
            # ll_to_hl returns 0 or more events
            # highlevel looks like: 
            # {type: "alphabloc", age: 42, duration: 300}
            current_node = add_event(current_node, highlevel)
\end{lstlisting}
\caption{Simple ESS Aggregation Algorithm.}\label{lst_simple}
\end{lstfloat}

\subsubsection{Optimizing Data Preprocessing\texorpdfstring{~\faBarChart}{}}
First, as shown in \autoref{fig_parcoursvisArchitecture}, ParcoursVis relies on pre-parsed data stored in binary format, avoiding the expensive step of parsing CSV data at each program start while ensuring all events per patient are sorted chronologically and that patients are \emph{shuffled} to accelerate the aggregation convergence. 
\am{This last step is useless in non-progressive systems but essential to optimize convergence and stability in PVA \faSpinner~\cite{PDABook}.}

Running the preprocessing step takes three minutes on our back end for 50 million events stored in CSV format. 
It is done once per dataset and generates the binary files that will be \jdf{loaded} when ParcoursVis's back end starts.
When the non-progressive program loads the CSV files in a time linear to the file size (\autoref{lst_simple} line 2), the progressive back end loads the binary files in negligible time using \emph{memory mapping} (\autoref{lst_aggregation} lines 3--4). \jdf{It is an operating system mechanism that places a local file into main memory instantly from the running program's perspective.}

\begin{lstfloat}
\begin{lstlisting}[language={Python},basicstyle=\ttfamily\small,escapechar=\%,morekeywords={\@dataclass, parallel}]
def ESS(directoryname):
    chunk_size = 100,000  # tuned to processing speed
    events = mmap_events(directoryname, "events")
    patients = mmap_patients(directoryname, "patients")
    full_tree = Node(type="root")
    for patients_chunk in split_by(patients, chunck_size):
        tree = parallel_aggregate(chunk, events)
        full_tree.merge(tree)
        render(root_tree)

@dataclass class Node:
    # No change from definition in %\autoref{lst_simple} line \autoref{lst_simple_node}%
    def merge(self, other):
        self.count += other.count
        self.agedist.update(other.agedist)
        self.durationdist += other.durationdist
        for (event, node) in other.children.items():
            if event in self.children:
                self.children[event].merge(node)
            else:
                self.children[event] = node

def parallel_aggregate(patients_chunk, events):
    thread_patients = split(patients, threadcount)
    trees = [None] * threadcount  # tree per thread
    parallel for thread in range(threadcount)
        tree = aggregate(thread_patients[thread], events)
        trees[thread] = tree
    for tree in trees[1:]:  # merge all in the first one
        trees[0].merge(tree)
    return trees[0]
\end{lstlisting}
\caption{PVA Parallelized ESS aggregation Algorithm. The type \texttt{Node} and function \texttt{aggregate} remain the same.}\label{lst_aggregation}
\end{lstfloat}

\subsubsection{Optimizing Aggregation\texorpdfstring{~\faBarChart~\faSpinner~$\rightrightarrows$}{}}
\label{optimize_aggreg}

Our aggregation algorithm (called \texttt{parallel\_aggregate} in \autoref{lst_aggregation} line 23) is similar to \texttt{aggregate} in \autoref{lst_simple} but rewritten for speed. It keeps the complexity linear with the total number of events~\faBarChart~but is much faster and guarantees a controlled latency. It is parallelized~$\rightrightarrows$ and, to become progressive, works by chunk~\faSpinner, split in line 6. \am{In the following, we describe our strategy based on the memory layout, data structures, data shuffling, and  parallelization.}

\faBarChart~The first optimization relies on the memory layout, which organizes the low-level event sequences in consecutive memory positions so the aggregation algorithm can access them sequentially. Modern hardware strongly optimizes this access pattern by prefetching memory caches along with sequential access~\cite{memory}.
The aggregation process takes a table of low-level events indexed per patient and outputs a prefix tree meant to be rendered.
In EventFlow, transforming low-level events into high-level ones is expensive due to the rich set of possible transformations supported, as shown in the \hyperref[sec_compareEventFlow]{Supplemental Material}. 
\jdf{In ParcoursVis, we} write the function \texttt{ll\_to\_hl} (low level to high level) directly in C++ for two reasons: expressive power and optimization.
\am{As explained in \autoref{sec_aggregation}, the aggregation functions are specific to treatments and use cases (\eg, non-cancerous prostate adenoma, used as an example here), so we hard-code the rules, \eg, the definition of what is an ``interruption'' synthetic event in a specific treatment, and make them follow state-of-the-art recommendations.}
\jdf{Still, we} allow the analysts to change some parameters within meaningful limits during the exploration \am{(see \autoref{fig_controlpanel_b})}.
ParcoursVis rules are more complex than what EventFlow can express, yet rather simple using a regular programming language like C++.
Our aggregation rules implementation is 190 lines long. 
Adapting it to another scenario, described in \autoref{sec_tasks}, took a few days and had a similar length.


\faSpinner~Second, for the progressive aspect, 
the main loop performs the tree aggregation along with the progressive calculations of the distributions. 
As shown in \autoref{lst_aggregation} line 6, the back end partially aggregates the tree with a certain number of sequences (a \emph{chunk}) and sends it to the front end (function \texttt{render} in \autoref{lst_aggregation} line 8). When the function returns, meaning the user interface received the data, it continues the aggregation for the next chunk.
To determine the \verb|chunk_size| parameter for a given iteration (kept constant in the pseudo-code of \autoref{lst_aggregation}), we measure the average speed (patients per second) of the last six iterations and multiply it by the desired latency (typically 2\comma s).
We start with the pessimistic chunk size of 100,000 patients and converge to a chunk size adapted to the network latency and performance of the back end.

We rely on the patients being shuffled \am{for faster convergence \jdf{of node frequency and average duration,} and to overcome order bias in the data~\cite{PDABook:3}.
Keeping patients in chronological order would initially reveal pathways that followed older medical protocols and delay the visualization of recent ones, biasing the early overview snapshots.}

\jdf{We optimize the resulting tree to the rendering client with \am{three} transformations: }
\jdf{(1) data structures stored in each node optimized for fast updates are transformed to be easily processed by the visualizations, \eg, the list of event durations is transformed into a histogram}, (2) small nodes below a specified threshold of patients (10--50) are trimmed, \am{and (3) node data not used by the tree layout \jdf{or the overview screen} are not included to minimize the transferred tree size}.
In our applications, the final tree has around 2\comma k nodes; \am{the trimmed version} can be transferred in a few milliseconds to the front end through a web socket.

\am{As we keep the computed tree in memory, we distinguish between two kinds of data transfer: always transfered, the trimmed tree and basic attributes to be visualized as distributions, and on-demand node-level details which are only sent when the user explicitly requests them by clicking a node.
This separation minimizes bandwidth usage and ensures low latency, even in environments with poor connectivity (e.g., hospitals with congested WiFi). With our strategy, web transfer latency only incurs when the user asks for details on demand, typically on a few nodes, and is low even when nodes contain rich aggregated data.}

$\rightrightarrows$ \autoref{lst_aggregation} Line 23 shows the parallelization of the algorithm.
It is ``embarrassingly parallel'' except for merging the resulting trees in lines 13-21, which could slow down the process. 
\autoref{sec_eval} evaluates the scalability of our aggregation algorithm \am{and shows the merging time is negligible}. 

\jdf{\faBarChart~Progressively updating the tree and merging nodes is fast because we use data structures (not specific to PVA) that are \emph{efficient to update} during sequence aggregation, and \emph{efficient to merge} during node merging.
}

\subsubsection{Progressive Rendering\texorpdfstring{~\faSpinner}{}}

Rendering is done on the front end in two passes: first, it preprocesses the tree according to user-specified parameters, such as the maximum tree depth and event filters, generating a rendering tree.
Then, it lays out the \IT in SVG format that eventually appears on the screen.
The first step rewrites the aggregated tree into a new one when only simple transforms are involved, saving back-end heavy work. Typically, to filter out node types or apply a hierarchy transform (explained in~\autoref{sec_controlpanel}).
This allows for a smoother user experience since a tree is usually composed of a few thousand nodes that are fast to process compared to the dataset that can contain millions of events. 

To prevent the progressive algorithm from flooding the rendering, the front end notifies the back end to continue the aggregation process once it has fully displayed an updated tree. 
When the user changes any of the parameters used for aggregation, the aggregation algorithm restarts from the beginning, iteratively updating the rendering, as shown by the arrows at the right of \autoref{fig_parcoursvisArchitecture}. 

\subsubsection{Stabilizing Rendering\texorpdfstring{~\faSpinner}{}}\label{sec_hysteresis}


\begin{lstfloat}
    \begin{lstlisting}[language={Python},basicstyle=\ttfamily\small,escapechar=\@]
def HysteresisSort(oldNode, newNode, inertia=20.0/1080):
  '''Sort nodes with a hysteresis condition
  oldNode: The node at the previous iteration
  newNode: The node at the current iteration.
  inertia: The hysteresis inertia
  return newNode with children sorted in descending order with an inertia'''
  if oldNode is None: # No previous iteration -> quick sort
    return sorted(newNode.children, key=lambda node: node.count,
                  reverse=True)
  # Apply a bubble sort with hysteresis to fulfill @\autoref{eq_hysteresisCondition}@.
  hasPermuted = True
  while hasPermuted:
    hasPermuted     = False
    minValue        = newNode.children[0].count
    lastMinValueIDx = 0
    for i in range(1, len(newNode.children)):
      # If @\autoref{eq_hysteresisCondition}@ is not fulfilled -> Reorder the nodes
      if minValue < newNode.children[i].count and \
         (newNode.children[i].count - minValue)/newNode.count > inertia:
        for j in range(i, lastMinValueIDx, -1):
          (newNode.children[j-1], newNode.children[j]) = \
            (newNode.children[j],   newNode.children[j-1])
        lastMinValueIDx += 1 # lastMinValue moved
        hasPermuted = True
      elif minValue >= newNode.children[i].count: 
        # using '>=' minimizes the permutations
        minValue = newNode.children[i].count
        lastMinValueIDx = i
  return newNode
    \end{lstlisting}
    \caption{The \emph{Hysteresis Sort} algorithm in Python.}\label{lst_hysteresis}
  \end{lstfloat}

The visualization of the ParcoursVis's aggregated tree is an \IT; it is laid out horizontally, node width encoding the average node duration and node height encoding the frequency, sorted to help compare the sibling frequency distribution per node.
The stability of the aggregated visualization between iterations mainly depends on two node parameters: the \emph{width} and the \emph{order} of siblings.
Changing the node width across updates can slightly shift subtrees, but changing sibling order can strongly change the overall layout, producing instability interpreted as uncertainty by users. We want to avoid this instability if it is only an artifact of the progressive algorithm.

With our progressive aggregation algorithm, nodes with close frequencies could switch positions between iterations, \eg, two nodes containing almost 20\% of the sequences each could swap back and forth across iterations due to slight variations in the distribution of patients (\eg, 19.9\% vs.\ 20.1\% and vice versa). We designed a \emph{Hysteresis Sort} algorithm (\autoref{lst_hysteresis}) to prevent these changes when the frequency difference is irrelevant visually. It is specifically designed for the progressive visualization of trees that rely on sorting the siblings by size.

Our \emph{Hysteresis Sort} sorts children perfectly at the first iteration.
For each new iteration, when the size of two adjacent nodes is almost the same, we consider the two sizes as equivalent and maintain their previous order, as shown in \autoref{lst_hysteresis}. 
Almost the same means the frequencies differ by a small number $\epsilon$; we call it the \emph{inertia}.
Our algorithm ensures that the nodes are almost sorted with an error bounded by the inertia, according to the following equation:
\begin{equation}\label{eq_hysteresisCondition}
    \forall \{i, j\}, i<j \implies \textit{freq}(\textit{child}_j)-\textit{freq}(\textit{child}_i) \le\epsilon
\end{equation}
We sort children in descending order, bottom to top, as we want new children created between two progressive iterations to be inserted at the end of the list of siblings (on top).
This is because, statistically, rare nodes appear later than frequent nodes in the progression.
Our algorithm is a variation of bubble sort. Since \autoref{eq_hysteresisCondition} does not fulfill the triangular inequality (if $a-b \leq \epsilon$ and $b-c \leq\epsilon$, we cannot be sure that $a-c \leq \epsilon$), we are forced to verify the whole list again when children are moved, using the variable \texttt{hasPermuted} (\autoref{lst_hysteresis} line 24). The algorithm has a worst-case complexity of $O(n^2)$, but since the children are mostly sorted, its actual complexity is quasi-linear, and the number of children is usually small \am{(ten in our prostate adenoma use case)}.
We choose a default inertia small enough to be unnoticeable visually, except at the first level, where it represents 20 pixels in height on a 1080\space p monitor. Below this value, two nodes are considered equivalent in size. Users can change inertia interactively if desired.


\section{Evaluation}\label{sec_eval}
\am{One of the primary goals of ParcoursVis is to ensure performance and scalability for the exploration of large-scale patient pathways, visualizing billions of aggregated events while maintaining interactivity. Rather than reinventing prior approaches, ParcoursVis builds upon EventFlow~\cite{Monroe:2013_TVCG}, extending its principles to operate at a much larger scale. Our evaluations confirm that we achieve comparable levels of efficiency and usability, even as we scale to data sizes several orders of magnitude larger. \jdf{We report metrics on scalability: latency, completion time, and stability.}
Accordingly, we report the time required for partial results to become sufficient for accurate decision-making in basic tasks. We provide all data analyses via \href{https://osf.io/xy2ev/?view_only=2853b8631e004955a9819108d41cdf53}{an osf.io repository}. \jdf{We also report on qualitative users' feedback.}
}

\subsection{Data}

ParcoursVis reads a dataset of low-level events, translates it into high-level ones, and builds its aggregated representation: the prefix tree.
Our evaluation relies on our original synthetic dataset of $2M$ patients treated for non-cancerous prostate adenoma. It uses five low-level event types---Alphabloc, Phyto, Fivealpha, Surgery, and Death---and translates them into ten high-level event types grouped into four higher-level categories, shown in \autoref{tab_events}.

\begin{table}
\begin{subtable}[t]{0.49\textwidth}
\begin{small}
\begin{tabular}[t]{llr}
\toprule
\textbf{Type} & \textbf{Super-type} & \textbf{\# of events} \\
\midrule
Alphabloc & Treatment & 1,098,585\\
Phyto & Treatment & 530,402\\
Fivealpha & Treatment & 192,127\\
Alphabloc Phyto & Treatment & 265,312\\
Fivealpha Alphabloc & Treatment & 161,166\\
Fivealpha Alphabloc Phyto & Treatment & 26,928\\
Interruption & Interruption & 123,465\\
No Treatment & Interruption & 111,188\\
Surgery & Surgery & 208,907\\
Death & Death & 1,738\\
\bottomrule
\end{tabular}
\end{small}
\caption{The ten event types used in our example use case.}
\label{tab_events}
\end{subtable}
\hspace{\fill}
\begin{subtable}[t]{0.49\textwidth}
\flushright
\begin{small}
\begin{tabular}[t]{rrrrr}
\toprule
\textbf{Patients} & \textbf{Low-Level} & \textbf{High-Level} & \textbf{Pathways} & \textbf{Nodes}\\
\midrule
 \textit{2M} &  \textit{50,640,284} &  \textit{2,719,818} & \textit{948} & \textit{1,122}\\
 4\space{M} &  73,621,135 &  5,441,380 & 926 & 1,109\\
 6\space{M} & 110,526,755 &  8,164,831 & 948 & 1,134\\
 8\space{M} & 147,300,619 & 10,884,766 & 966 & 1,155\\
10\space{M} & 184,161,763 & 13,607,620 & 972 & 1,161\\
\bottomrule
\end{tabular}
\end{small}
\caption{Dataset sizes used in our evaluation.}\label{tab_sizes}
\end{subtable}
\caption{Event types and numbers used in our experiments.}
\end{table}

To measure the scalability of ParcoursVis, we generated four synthetic datasets with $\textit{size} \in \{4\space \textrm{M}, 6\space \textrm{M}, 8\space \textrm{M}, 10\space \textrm{M}\}$ patients using the strategy described in the \hyperref[sec_syntheticData]{supplementary material}; their characteristics are listed in \autoref{tab_sizes}.
The aggregated tree characteristics of the original 2\space{M} patients dataset are in the first line.

\begin{figure}[ht]
    \centering
    \begin{subfigure}[b]{0.7\linewidth}
        \includegraphics[width=\linewidth]{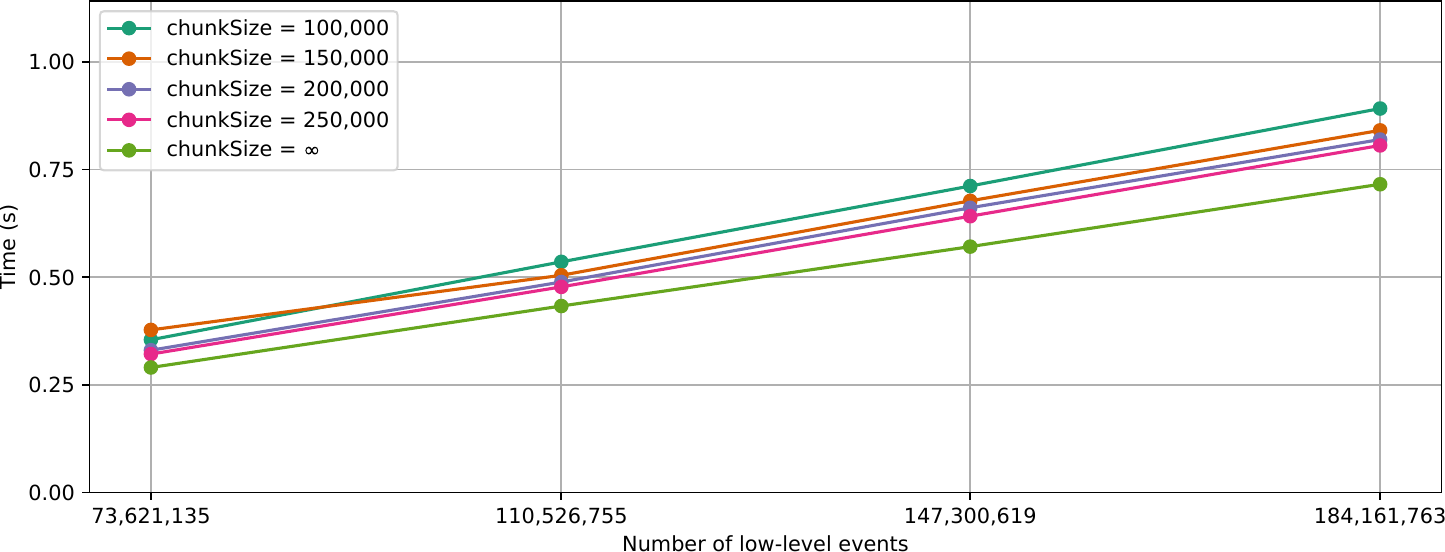}
        \caption{Total computation time per size with five chunk sizes using 6 threads. $\textit{chunkSize} = \infty$ means non-progressive.}
        \label{fig_resultsScalability_chunk}
    \end{subfigure}
    
    \begin{subfigure}[b]{0.7\linewidth}
        \includegraphics[width=\linewidth]{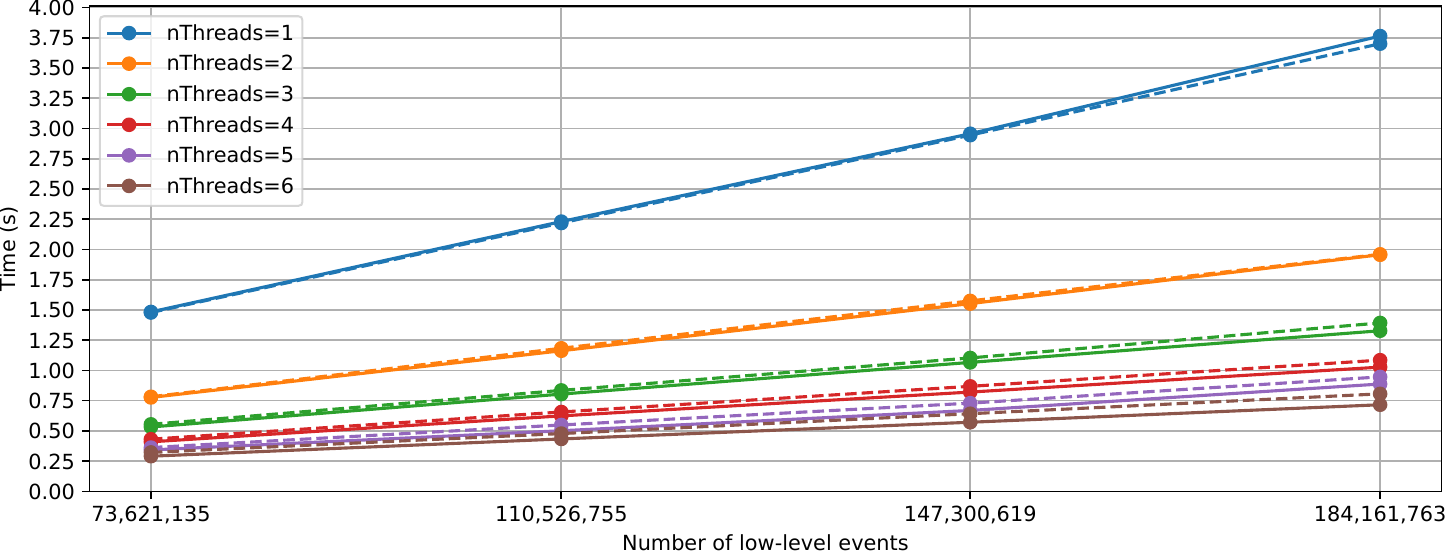}
        \caption{Total computation time per size compared by \# of threads in progressive (dotted) and non-progressive (plain) settings.}
        \label{fig_resultsThreadsTimeNbEvents}
    \end{subfigure}
    \caption{Scalability comparison across chunk sizes (a) and thread counts (b).}
    \label{fig_combinedScalability}
    \Description{TODO}
\end{figure}

Before measuring the stability of ParcoursVis, we analyzed the time to aggregate the dataset with our progressive algorithm (\ie, chunk by chunk) compared to treating the whole dataset at once, not progressively. This \emph{competitive analysis} is standard for online algorithms~\cite{online-survey}.
The reported completion times do not consider the rendering of the tree nor the possible network delay of a back-end/front-end architecture.

\subsection{Aggregation}\label{sec_competitiveAnalysis}

We expect our algorithm (\autoref{lst_aggregation}) to be linear in the number of events and threads. We report the computation times it takes to aggregate them both progressively and non-progressively. Instead of aggregating patients over a quantum of time, we aggregate with a fixed number of sequences for better control and reproducibility~\cite{reproducibility}. We vary $\textit{chunkSize}\in\{100\space \textrm{k}, 150\space \textrm{k}, 200\space \textrm{k}, 250\space \textrm{k}\}$ sequences. 
As we expect our algorithm to be linear in the number of low-level events, we use that metric as the size of the databases in our evaluation. We ran our evaluations on an Intel(R) Core(TM) i7-8700K CPU @ 3.70GHz that has 6 cores and 12 logical threads. We disabled simultaneous multithreading (SMT) to enforce that each thread is associated with a physical core. We measured the time using a monotonic wall-clock.

\subsubsection{Results}

It takes from 3.75\space{s} with one thread (2.67\space{M} patients per second) to 0.75\space{s} with six threads (13\space{M} patients per second) to process our biggest synthetic dataset of 10\space{M} patients in a non-progressive environment (\autoref{fig_resultsScalability_chunk}).
\am{This result, which is below the 10\space{s} threshold introduced in \autoref{sec_PVA}, should still be interpreted with caution. If the back-end machine is heavily loaded or if the network experiences slowdowns, the response time may become unstable and quickly exceed the 10\space{s} threshold. Additionally, datasets that cover a large country or involve a more complex pathology, encompassing larger patient cohorts, may also lead to exceeding this limit.}
The progressive environment yields similar results but with a time overhead that depends on the \textit{chunkSize} parameter. \autoref{fig_resultsScalability_chunk} shows that the computation time decreases when the chunk size increases. \autoref{fig_resultsThreadsTimeNbEvents} also shows that the overhead increases with the number of threads. For the mono-thread configuration, the overhead is negligible.
We explain this overhead by the number of tree merges the algorithm does (see \autoref{lst_aggregation} line 30), which depends on the average and the number of threads as follows: 
$\textit{merges} = (\textit{nbThreads}-1)\times\textit{nbPatients}/\textit{chunkSize}$.
For \textit{chunkSize}=100\space k, \textit{nbPatients}=10\space M, and \textit{nbThreads}=6, we merge the trees with $\approx 1,100$ nodes 500 times.
For the best-case scenario of \textit{chunkSize}=250\space k and \textit{nbThreads}=6, each progressive iteration takes about 19.8\space ms (median) to complete (12.6\space M patients per second).


\autoref{fig_resultsThreadsTimeNbEvents} shows that, for all numbers of threads, the computation time of our algorithm remains linear in the number of low-level events for both progressive and non-progressive environments.
Plotting the total computation time per size compared to the number of threads, we find that our algorithm's running time is inversely proportional to the number of threads. 
Looking at the computational speed per size compared to mono-thread computations, we find that the speed increases linearly with the number of cores. A linear least-squares regression gives a slope $\approx 0.83$ and $R^2 > 0.99$. A one-way ANOVA on the linear regression predictions gives a $\textit{p-value} > 0.99$; the slope coefficients for all the chunk sizes are the same.

We conclude that the runtime of our progressive and non-progressive algorithms is inversely linear to the number of cores and linear with the number of low-level events. Tree merge time is negligible.

\subsection{Stability}\label{sec_stability}

We now focus on the stability and convergence of ParcoursVis over the progressive iterations. 
%
%
We compare our \emph{Hysteresis Sort}, described in \autoref{sec_hysteresis}, with the regular sort (called \emph{RSort} in the remaining) that EventFlow uses, \ie, sorting nodes based on frequency.
We configured \emph{Hysteresis Sort} with $\textit{inertia}=\frac{20}{1080}\approx0.02$ by default, which corresponds to 20 pixels for the root level of our visualization on a standard 1080\space p monitor, and below one pixel for the deeper tree levels.

We considered two external factors affecting the ordering of nodes. First, we avoid what Monroe \etal~\cite{Monroe:2013_TVCG} call \emph{confetti} visualization by only showing nodes representing more than \emph{MinSize} number of patients.
We evaluate ParcoursVis along the \emph{MinSize} variable: we output a node and its children only if $\textit{size}(\textit{node}) \ge \textit{MinSize}, \textit{MinSize} \in \{0, 25, 50\}$.
The second factor impacting the stability of the tree is the convergence of the intermediate progressive results, which is sensitive to the number of sequences each iteration processes; larger chunks will lead to faster convergence. The number of intermediate steps depends on the CPU speed, machine load, and the time quantum, which are hard to control. We, therefore, vary the $\textit{chunkSize} \in \{100\space \textrm{k}, 150\space \textrm{k}, 200\space \textrm{k}, 250\space \textrm{k}\}$ to control the number of patients processed per iteration to evaluate when nodes stabilize while ensuring reproducibility for the evaluation.

\subsubsection{Metrics of Stability}

For \emph{Hysteresis Sort} and \emph{RSort}, we report two metrics for stability:

\noindent\textbf{Stability by Depth} A tree depth is considered stable at iteration $i$ if, below that depth, there is no change of order in the nodes generated from $\textit{iteration}=i$ to the final result.

\noindent\textbf{Stability Per Node} A node is considered stable if its parent is stable, if this node and its siblings have the same rank in their parent's list from $\textit{iteration}=i$ to the final result. The root is always considered stable.
If a node is not created at a given iteration, we consider that it is placed at the end (lowest frequency, last rank) of its sibling list.

We report the lowest possible $\textit{iteration}=i$ for both metrics.
%
We emphasize that since we compare nodes from $\textit{iteration}=i$ to the final result, our report on stability is similar to convergence.

\begin{figure}[t]
    \centering
    \begin{subfigure}[b]{0.7\linewidth}
        \includegraphics[width=\linewidth]{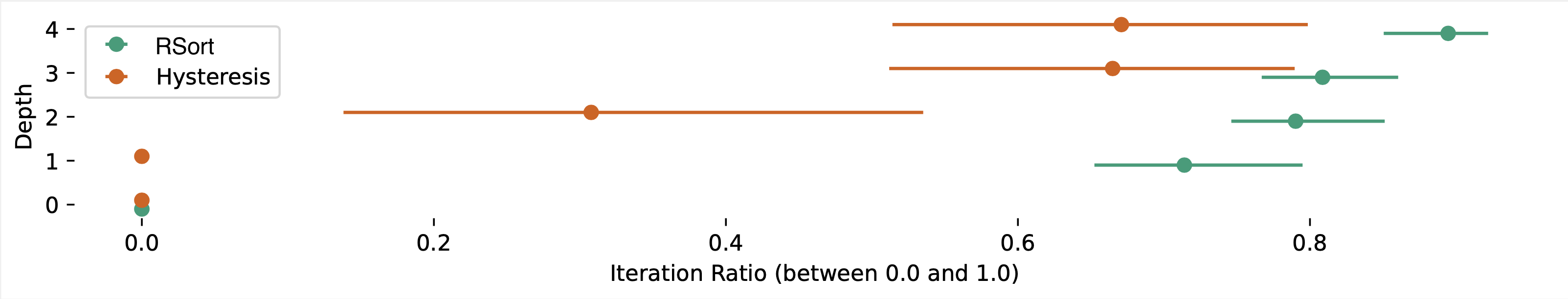}
        \caption{Trees' stabilization speed for \emph{Hysteresis Sort} and \emph{RSort} by depth.\newline Iteration ratio $[0,1]$ indicates dataset processing progress.}
        \label{fig_stability}
    \end{subfigure}
    
    \begin{subfigure}[b]{0.7\linewidth}
        \includegraphics[width=\linewidth]{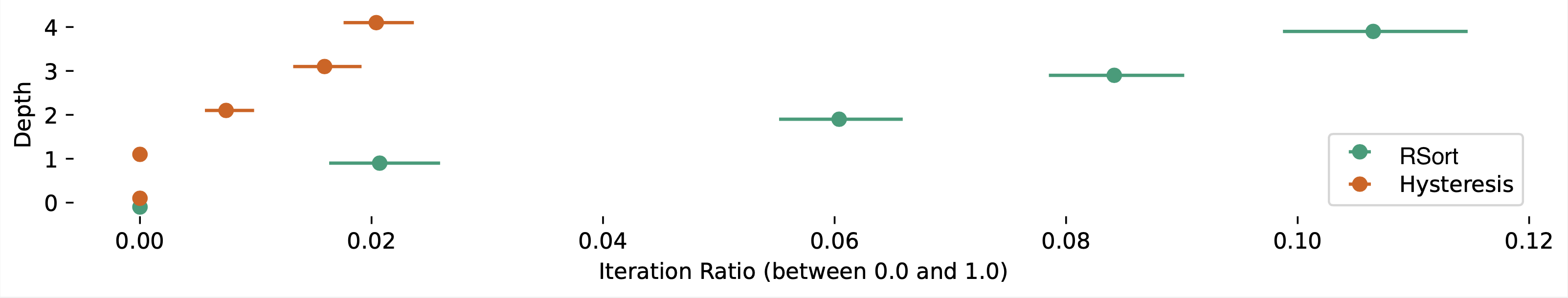}
        \caption{Stabilization speed of the nodes using \emph{Hysteresis Sort} and \emph{RSort}  by depth.}
        \label{fig_stabilityCINodes}
    \end{subfigure}


    \begin{subfigure}[b]{0.7\linewidth}
        \includegraphics[width=\linewidth]{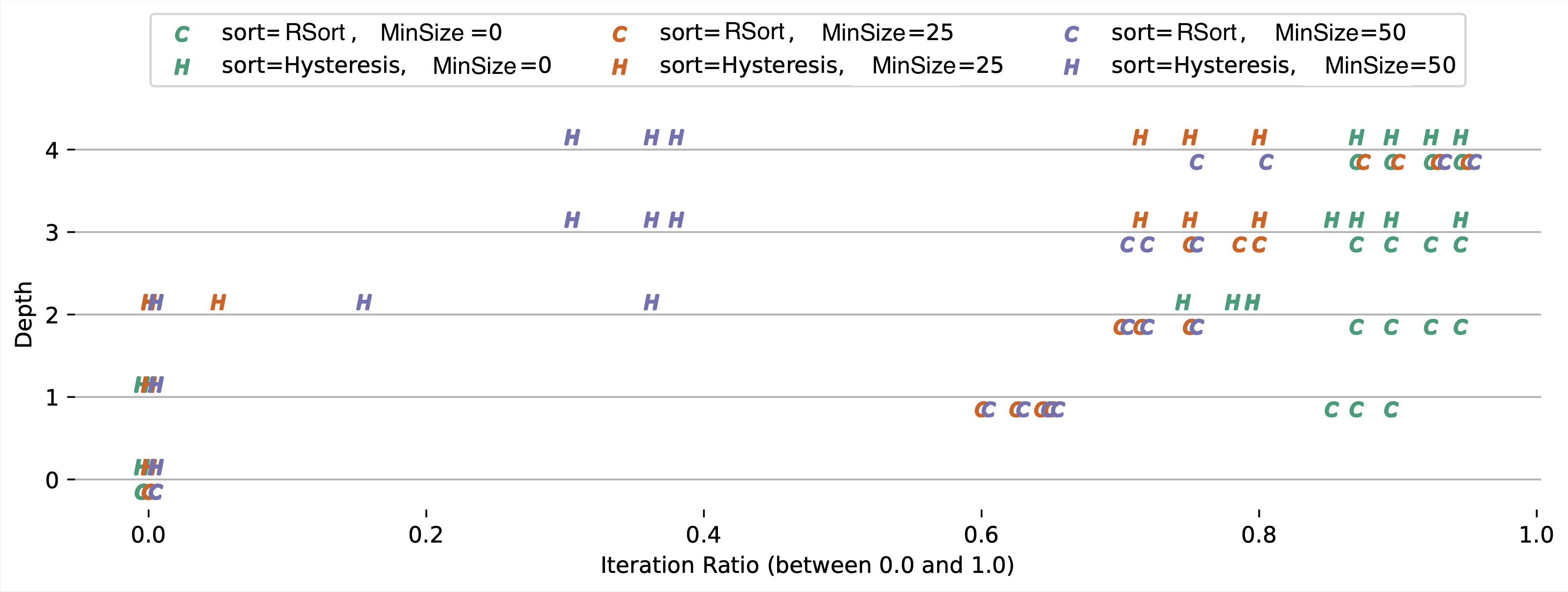}
        \caption{Stability of the tree by depth and \textit{MinSize}. Each category shows 4 \textit{chunkSize} values. Points are jittered on the x-axis to avoid overlap. 
        Larger \textit{MinSize} leads to earlier stabilization, especially in \emph{Hysteresis Sort}.}
        \label{fig_perThreshold}
    \end{subfigure}

    \caption{Stability analysis of \emph{Hysteresis Sort} and \emph{RSort}: (a–b) stabilization results by depth and node count, and (c) effect of \textit{MinSize} on overall tree stability.}
    \label{fig_stabilityCombined}
    \Description{TODO}
\end{figure}

\subsubsection{Results}
Our evaluation shows that all nodes have a similar frequency from the first iteration to the last, regardless of \emph{MinSize} and \emph{chunkSize}, meaning that the frequency distribution of nodes converges fast for datasets like ours. As we shuffled the datasets, this result is expected for high values of \emph{chunkSize}.

We then look at how fast \emph{Hysteresis Sort} and \emph{RSort} stabilize in depth and relative position (\autoref{fig_stability} and \autoref{fig_stabilityCINodes}). 
We normalized the iteration value to $[0, 1]$ to compare the evaluations with different \emph{chunkSize}.
The distribution of the number of stabilized nodes and depths by iteration value is not normal.
Thus, we cannot rely on parametric statistical tests. Following recommendations of~\cite{Dragicevic:2016_FSC, Amrhein:2019_SSS, Besancon:2019_CPDI}, we instead rely on estimation techniques with effect-size and confidence intervals (CIs). 
We bootstrap all the evaluation results per sorting strategy and per depth, resulting in $\textit{size}(\textit{MinSize}) \times \textit{size}(\textit{chunkSize})$ different values per bootstrap. 
 Depth starts from 0, which corresponds to the children of the root, as the root is, in our case, an abstract object that we do not visualize in ParcoursVis.

\autoref{fig_stability} shows that the prefix tree is stable after the progressive algorithm processed $80\%$ (at the right edge of the 95\% CI) of the dataset at the maximum tested $\textit{depth}=4$ using the \emph{Hysteresis Sort}, compared to approximately 90\% using \emph{RSort}. 
Our results
(\autoref{fig_stabilityCINodes}) show statistically significant differences that \emph{Hysteresis Sort} and \emph{RSort} start to diverge at \emph{depth=1}, with a large effect size for $\textit{depth} \in \{1,2\}$. Because deeper depths depend on shallower ones, the results of \emph{RSort} for $\textit{depth}=1$ strongly impact the deeper results. This seems less the case for \emph{Hysteresis Sort} (\autoref{fig_stability}). 

We then looked at the stability per node, categorized by their depths. \autoref{fig_stabilityCINodes} shows strong evidence that \emph{Hysteresis Sort} outperforms \emph{RSort} at stabilizing nodes sooner for all depths. In addition, we looked at the stability of nodes categorized by their frequencies. We computed, per node, the difference in stability of \emph{Hysteresis Sort} compared to \emph{RSort} for the evaluation, set with $\textit{MinSize}=0$ and $\textit{chunkSize}=100\space \textrm{k}$, which are the minimum values we tested. We see that \emph{Hysteresis Sort} and \emph{RSort} behave the same for most nodes, which we expected, as most nodes do not have siblings with similar frequencies (see $\textit{iteration}=0$). Second, the sorting strategy mostly impacts nodes with low frequency (lower than 200, \ie, representing less than 1\% of our population). Third, while \emph{Hysteresis Sort} strongly improves the stability of most of the remaining nodes, it can degrade the stability of some small nodes for a few iterations.
Overall, we conclude on the effectiveness of \emph{Hysteresis Sort}  compared to \emph{RSort}. \am{With our datasets and standard setting of \textit{MinSize}=50, the tree is completely stable when 40\% of the dataset has been processed (\autoref{fig_perThreshold}).}

Finally, we wanted to see if the variables \textit{MinSize} and \textit{chunkSize} have an effect on the stability. %
\autoref{fig_perThreshold} suggests that \textit{MinSize} strongly impacts the stability of the positions of the nodes by removing small changes. Because the dataset is shuffled, large nodes tend to stabilize sooner than smaller ones, explaining the results. However, this figure shows that \emph{MinSize} has a stronger impact on the \emph{Hysteresis Sort} than on the \emph{RSort}.
We do not find any effect of the \textit{chunkSize} variable on the stability of nodes. 

\subsection{User Feedback}
\am{We collected user feedback from both use cases introduced in \autoref{sec_tasks}, enabling participants to visualize datasets with an unprecedented level of scalability with ParcoursVis. We provide usability insights, focusing on its progressive aspect when used by experts. 
Participants were able to perform tasks comparable to those in EventFlow systems with no perceived latency. This study has been approved by our organization's Ethical Committee.}

\am{Experts were excited to be able to explore such large datasets to eventually find unexpected cases. We interviewed a referring doctor to explore the non-cancerous prostate adenoma dataset and received positive feedback.
He reported that his wife thought he had a new video game because he was spending hours exploring the data interactively. 
He also collected many insights and surprising findings that need inquiry now. For example, he thought that surgery would cure the patients when, in reality, many patients ($18\%$) need to resume treatment and sometimes undergo another surgery. He also formulated hypotheses that require validation through further examination of patients' cases (\eg, long drug treatments on severe cases may weaken the urinary tract).}

\am{We interviewed six additional healthcare professionals on the ED dataset: four emergency physicians (two with extensive experience in ED operations and two focused on improving organizational efficiency), and two clinical research data analysts. We also interviewed a researcher in computer science focused on high-fidelity simulation of patient flow in emergency departments. During the study, we used data from three hospitals for a total of 476,109 visitors for reasons of data quality and comparability. We asked them to complete some of the tasks listed in \autoref{sec_tasks} using a think-aloud approach, followed by a semi-structured interview; all sessions were audio-recorded.}

\jdf{They all succeeded in performing the tasks.}
\emph{None of our participants mentioned any latency issues, not even performance issues. They were able to focus on their professional questions.} 

\jdf{Additionally, we used ParcoursVis to answer several data-related questions during project meetings. For example, we were able to quickly characterize and select important pathways to optimize, such as the most frequent ones, the most time-consuming ones, and the most ``undesirable'' ones, \eg, revisiting the ICU, or the emergency service in a short time.}

\section{Discussion and Future Work}

\subsection{Performance Analysis}
The analysis described in \autoref{sec_competitiveAnalysis} validates the linear complexity of our aggregation algorithm, both in the number of threads and low-level events. We aggregate 10 million synthetic patients ($\approx 180\space \textrm{M}$ low-level events) on our dedicated computer in less than one second. ParcoursVis is meant to be used on a server shared by multiple users; aggregating sequences in the wild takes longer. Moreover, our evaluation relies on short sequences on which we applied a small number of filters. Longer sequences and more filters cause slowdowns. PVA solves these issues with almost no overhead.

We filter small nodes to avoid confetti visualizations, but users can configure it. In our evaluation using $\textit{MinSize}=50$ (\ie, 0.0025\% of the two million patients dataset), the view aggregating the dataset stabilizes at around 40\% of the computation time for the deepest nodes using \emph{Hysteresis Sort}, 25\% sooner than \emph{RSort} (\autoref{fig_perThreshold}). The \emph{MinSize} variable thus can strongly increase the speed at which users can work on their data if they are not interested in small nodes.

\subsection{Stability}
We evaluated the stability of aggregated trees for one use case with an inertia of $20/1080$. For that configuration, we show \emph{Hysteresis Sort} is better than \emph{RSort}. We believe this generalizes to other datasets, although the stabilization times may vary depending on the distribution of events.
Our \emph{Hysteresis Sort} algorithm relies on a modified version of the bubble sort algorithm with a theoretical worst-case quadratic complexity. 
In practice, our sort time is linear because we use a limited number of types to be able to assign them to color effectively (as explained in \autoref{sec_scalability}).
The scalability of ParcoursVis and ESS, in general, remains limited by the number of event types they can handle~\cite{Wang:2022_CGF}.



\subsection{Evaluation Criteria for Progressive Visualization}

Progressive systems are relatively new, and we need more evaluation criteria to assess their usability~\cite{PDABook:9}.
We did not consider comparing the usability of our system in its current state to a non-progressive system, given the latency we measured, which makes non-progressive systems completely unusable (measures are in the \hyperref[sec_compareEventFlow]{Supplementary Material}). During our expert interviews, participants were unaware that the tool used progressive algorithms and did not notice it; we consider this lack of mention a usability success for PVA. \emph{Yet, it is a new qualitative evaluation criterion. }


As explained in \autoref{sec_uncertainty}, we rely on stability as a proxy for convergence.
For update durations limited to tens of seconds, we believe instability is an acceptable proxy for uncertainty, and our expert users did not express any confusion or trust issues about it.
Longer update times may need explicit uncertainty visualization, though.

\subsection{Improving the Scalability of EHR Visualizations}



Enhancing the efficiency of data structures,  algorithms~\faBarChart~and using parallelism~$\rightrightarrows$ are crucial for scalability. Yet, relying solely on traditional optimizations also overlooks the potential for orders of magnitude improvements with PVA~\faSpinner~.
The transition to a progressive approach, while offering substantial benefits for achieving scalability, is nonetheless far from straightforward and relies on various implementation choices.
As demonstrated in this article, partitioning data into chunks and effectively managing aggregated cumulative data are crucial for optimizing performance at scale. 
Indeed, incorporating progressive computation in the main loop further enhances scalability and decouples interactivity from data size~\faSpinner~.
In the latest version of ParcoursVis, we also perform progressive computations on demand, such as the distributions of a large number of attributes associated with patients when a node is selected; this avoids slowing down the main aggregation loop while supporting a richer set of attributes.


Furthermore, effective uncertainty visualization depends on the convergence rate. When the convergence is longer than 30 seconds or so, uncertainty can be visualized on its own since users have time to read it and decide if what they see is good enough while the progression continues. For faster convergence, stability can become a proxy for quality with the caveat of avoiding spurious instability due to progressive algorithm artifacts, as described in \autoref{sec_stability}.





\subsection{Future Work}
Each new medical study requires new features, and we are planning to extend ParcoursVis to make it more generic and applicable to new use cases with little programming, providing a larger set of aggregation options as a library.

We also plan to support datasets updated continuously. That would require certain adjustments to the system, specifically running regular pre-processing of the updated database. 
ParcoursVis should be able to restart immediately when a new dataset is available, but it should also avoid interrupting ongoing explorations.

ParcoursVis is currently being applied to a new pathology, prostate cancer, using data from the SNDS~\cite{SNDS}. While we currently handle only a limited number of attributes per patient, the dataset provides thousands of attributes. To enable richer exploration, we aim to develop methods to manage these attributes dynamically and on demand, including the ability to display their distributions, scaling ParcoursVis to effectively support another effort: a large number of attributes and types. One way to achieve this, building on the idea introduced in \autoref{optimize_aggreg}, is to keep the patient identifiers in the aggregation tree and compute patient-level aggregations on the fly from the node patient list. We would rely on an external database such as DuckDB~\cite{DuckDB} to compute the distributions or aggregated values. This approach would scale ParcoursVis to handle not only a large number of patients but also a large variety of attributes and types, without overloading bandwidth or memory.


\section{Conclusion}
We presented ParcoursVis, a Progressive Visual Analytics tool we designed to explore patients' care pathways at scale.
We described its features, including the history navigation and comparison, which are novel for EHR visualization systems.
We described its progressive architecture, scalable aggregation algorithm, and its stabilizing sorting algorithm.
Using Richer \etal~\cite{Richer:2022_TVCG} scalability model, we applied quantitative evaluations showing the aggregation algorithm scales linearly with the number of events and inversely with the number of threads, always meeting interactive latency limits. We also showed that our Hysteresis sort improves the stability of the \IT visualization.
ParcoursVis can be used to explore patients' treatments and conditions aggregated at the scale of the largest countries to improve public health based on data.

For a dataset similar to our benchmark scaled to China’s population, non-progressive aggregation on our machine would take 27\space{s} to complete.
In contrast, our PVA architecture ensures updates in under 2\space{s}, with most tree nodes stabilizing within 10\space{s}, enabling exploration of datasets at the scale of the largest countries.



\jdf{Adapting visualizations to become progressive and scalable still needs experience and novel software architectures; this article explained how we addressed important issues.}
\jdf{Increasing the scalability of aggregated sequence visualization by three to five orders of magnitude, depending on the measure, allows addressing new sets of problems from larger medical data.}
By providing our system as open-source, we want to push this research field to a wider audience, relying on PVA to provide scalability and improve public health worldwide.


\ifanon
\else
\section*{Acknowledgments}

The authors wish to thank Catherine Plaisant for her valuable input, which mostly comes from her experience with the EventFlow project.
This work was supported in part by a grant from the Health Data-Hub, and from the \href{https://www.inria.fr/en/urge}{URGE AP-HP/Inria project}.

\fi

\bibliographystyle{ACM-Reference-Format}
\bibliography{sample-base}


\begin{thebibliography}{57}


\ifx \showCODEN    \undefined \def \showCODEN     #1{\unskip}     \fi
\ifx \showISBNx    \undefined \def \showISBNx     #1{\unskip}     \fi
\ifx \showISBNxiii \undefined \def \showISBNxiii  #1{\unskip}     \fi
\ifx \showISSN     \undefined \def \showISSN      #1{\unskip}     \fi
\ifx \showLCCN     \undefined \def \showLCCN      #1{\unskip}     \fi
\ifx \shownote     \undefined \def \shownote      #1{#1}          \fi
\ifx \showarticletitle \undefined \def \showarticletitle #1{#1}   \fi
\ifx \showURL      \undefined \def \showURL       {\relax}        \fi
\providecommand\bibfield[2]{#2}
\providecommand\bibinfo[2]{#2}
\providecommand\natexlab[1]{#1}
\providecommand\showeprint[2][]{arXiv:#2}

\bibitem[Amrhein et~al\mbox{.}(2019)]%
        {Amrhein:2019_SSS}
\bibfield{author}{\bibinfo{person}{Valentin Amrhein}, \bibinfo{person}{Sander
  Greenland}, {and} \bibinfo{person}{Blake McShane}.}
  \bibinfo{year}{2019}\natexlab{}.
\newblock \showarticletitle{Scientists rise up against statistical
  significance}.
\newblock \bibinfo{journal}{\emph{Nature}} \bibinfo{volume}{567},
  \bibinfo{number}{7748} (\bibinfo{year}{2019}), \bibinfo{pages}{305--307}.
\newblock
\href{https://doi.org/10.1038/d41586-019-00857-9}{doi:\nolinkurl{10.1038/d41586-019-00857-9}}


\bibitem[Angelini et~al\mbox{.}(2018)]%
        {Angelini:2018_Informatics}
\bibfield{author}{\bibinfo{person}{Marco Angelini}, \bibinfo{person}{Giuseppe
  Santucci}, \bibinfo{person}{Heidrun Schumann}, {and}
  \bibinfo{person}{Hans-Jörg Schulz}.} \bibinfo{year}{2018}\natexlab{}.
\newblock \showarticletitle{{A Review and Characterization of Progressive
  Visual Analytics}}.
\newblock \bibinfo{journal}{\emph{Informatics}} \bibinfo{volume}{5},
  \bibinfo{number}{3} (\bibinfo{year}{2018}), \bibinfo{pages}{31:1--31:27}.
\newblock
\showISSN{2227--9709}
\href{https://doi.org/10.3390/informatics5030031}{doi:\nolinkurl{10.3390/informatics5030031}}


\bibitem[Arleo et~al\mbox{.}(2024)]%
        {10.1093/jamia/ocae249}
\bibfield{author}{\bibinfo{person}{Alessio Arleo}, \bibinfo{person}{Annie~T
  Chen}, \bibinfo{person}{David Gotz}, \bibinfo{person}{Swaminathan
  Kandaswamy}, {and} \bibinfo{person}{Jürgen Bernard}.}
  \bibinfo{year}{2024}\natexlab{}.
\newblock \showarticletitle{Reflections on interactive visualization of
  electronic health records: past, present, future}.
\newblock \bibinfo{journal}{\emph{Journal of the American Medical Informatics
  Association}} \bibinfo{volume}{31}, \bibinfo{number}{11} (\bibinfo{date}{10}
  \bibinfo{year}{2024}), \bibinfo{pages}{2423--2428}.
\newblock
\showISSN{1527-974X}
\showeprint{https://academic.oup.com/jamia/article-pdf/31/11/2423/59813692/ocae249.pdf}
\href{https://doi.org/10.1093/jamia/ocae249}{doi:\nolinkurl{10.1093/jamia/ocae249}}


\bibitem[Bernard et~al\mbox{.}(2015)]%
        {bernard2015visual}
\bibfield{author}{\bibinfo{person}{J{\"u}rgen Bernard}, \bibinfo{person}{David
  Sessler}, \bibinfo{person}{Thorsten May}, \bibinfo{person}{Thorsten Schlomm},
  \bibinfo{person}{Dirk Pehrke}, {and} \bibinfo{person}{J{\"o}rn Kohlhammer}.}
  \bibinfo{year}{2015}\natexlab{}.
\newblock \showarticletitle{A visual-interactive system for prostate cancer
  cohort analysis}.
\newblock \bibinfo{journal}{\emph{{IEEE} Computer Graphics and Applications}}
  \bibinfo{volume}{35}, \bibinfo{number}{3} (\bibinfo{year}{2015}),
  \bibinfo{pages}{44--55}.
\newblock


\bibitem[Besan{\c c}on and Dragicevic(2019)]%
        {Besancon:2019_CPDI}
\bibfield{author}{\bibinfo{person}{Lonni Besan{\c c}on} {and}
  \bibinfo{person}{Pierre Dragicevic}.} \bibinfo{year}{2019}\natexlab{}.
\newblock \showarticletitle{{The Continued Prevalence of Dichotomous Inferences
  at CHI}}. In \bibinfo{booktitle}{\emph{Proc.\ CHI}}.
  \bibinfo{publisher}{ACM}, \bibinfo{address}{New York},
  \bibinfo{pages}{14:1--14:11}.
\newblock
\href{https://doi.org/10.1145/3290607.3310432}{doi:\nolinkurl{10.1145/3290607.3310432}}


\bibitem[Bjarnad\'ottir et~al\mbox{.}(2016)]%
        {Bjarnadottir:2016_PharmacoEconomics}
\bibfield{author}{\bibinfo{person}{Margr\'et~V Bjarnad\'ottir},
  \bibinfo{person}{Sana Malik}, \bibinfo{person}{Eberechukwu Onukwugha},
  \bibinfo{person}{Tanisha Gooden}, {and} \bibinfo{person}{Catherine
  Plaisant}.} \bibinfo{year}{2016}\natexlab{}.
\newblock \showarticletitle{{Understanding Adherence and Prescription Patterns
  Using Large-Scale Claims Data}}.
\newblock \bibinfo{journal}{\emph{PharmacoEconomics}}  \bibinfo{volume}{34}
  (\bibinfo{year}{2016}), \bibinfo{pages}{169--179}.
\newblock
Issue 2.
\href{https://doi.org/10.1007/s40273-015-0333-4}{doi:\nolinkurl{10.1007/s40273-015-0333-4}}


\bibitem[Borodin and El-Yaniv(1998)]%
        {online-survey}
\bibfield{author}{\bibinfo{person}{Allan Borodin} {and} \bibinfo{person}{Ran
  El-Yaniv}.} \bibinfo{year}{1998}\natexlab{}.
\newblock \bibinfo{booktitle}{\emph{{Online Computation and Competitive
  Analysis}}}.
\newblock \bibinfo{publisher}{Cambridge University Press},
  \bibinfo{address}{New York, NY, United States}.
\newblock
\showISBNx{0-521-56392-5}


\bibitem[Bostock et~al\mbox{.}(2011)]%
        {D3}
\bibfield{author}{\bibinfo{person}{Michael Bostock}, \bibinfo{person}{Vadim
  Ogievetsky}, {and} \bibinfo{person}{Jeffrey Heer}.}
  \bibinfo{year}{2011}\natexlab{}.
\newblock \showarticletitle{{D{\({^3}\)} Data-Driven Documents}}.
\newblock \bibinfo{journal}{\emph{{IEEE} Trans. Vis. Comput. Graph.}}
  \bibinfo{volume}{17}, \bibinfo{number}{12} (\bibinfo{year}{2011}),
  \bibinfo{pages}{2301--2309}.
\newblock
\href{https://doi.org/10.1109/TVCG.2011.185}{doi:\nolinkurl{10.1109/TVCG.2011.185}}


\bibitem[Brickell and Shmatikov(2008)]%
        {Brickell:2008:KDD}
\bibfield{author}{\bibinfo{person}{Justin Brickell} {and}
  \bibinfo{person}{Vitaly Shmatikov}.} \bibinfo{year}{2008}\natexlab{}.
\newblock \showarticletitle{{The Cost of Privacy: Destruction of Data-Mining
  Utility in Anonymized Data Publishing}}. In \bibinfo{booktitle}{\emph{Proc.\
  KDD}}. \bibinfo{publisher}{ACM}, \bibinfo{address}{New York},
  \bibinfo{pages}{70--–78}.
\newblock
\href{https://doi.org/10.1145/1401890.1401904}{doi:\nolinkurl{10.1145/1401890.1401904}}


\bibitem[Correll and Gleicher(2014)]%
        {Correll:2014_TVCG}
\bibfield{author}{\bibinfo{person}{Michael Correll} {and}
  \bibinfo{person}{Michael Gleicher}.} \bibinfo{year}{2014}\natexlab{}.
\newblock \showarticletitle{{Error Bars Considered Harmful: Exploring Alternate
  Encodings for Mean and Error}}.
\newblock \bibinfo{journal}{\emph{{IEEE} Trans. Vis. Comput. Graph.}}
  \bibinfo{volume}{20}, \bibinfo{number}{12} (\bibinfo{year}{2014}),
  \bibinfo{pages}{2142--2151}.
\newblock
\href{https://doi.org/10.1109/TVCG.2014.2346298}{doi:\nolinkurl{10.1109/TVCG.2014.2346298}}


\bibitem[Cutler et~al\mbox{.}(2020)]%
        {Cutler:2020_VIS}
\bibfield{author}{\bibinfo{person}{Zach Cutler}, \bibinfo{person}{Kiran
  Gadhave}, {and} \bibinfo{person}{Alexander Lex}.}
  \bibinfo{year}{2020}\natexlab{}.
\newblock \showarticletitle{{Trrack: A Library for Provenance-Tracking in
  Web-Based Visualizations}}. In \bibinfo{booktitle}{\emph{Proc.\ VIS}}.
  \bibinfo{publisher}{IEEE}, \bibinfo{address}{Los Alamitos},
  \bibinfo{pages}{116--120}.
\newblock
\href{https://doi.org/10.1109/VIS47514.2020.00030}{doi:\nolinkurl{10.1109/VIS47514.2020.00030}}


\bibitem[Dragicevic(2016)]%
        {Dragicevic:2016_FSC}
\bibfield{author}{\bibinfo{person}{Pierre Dragicevic}.}
  \bibinfo{year}{2016}\natexlab{}.
\newblock \showarticletitle{{Fair Statistical Communication in HCI}}.
\newblock In \bibinfo{booktitle}{\emph{{Modern Statistical Methods for HCI}}}.
  \bibinfo{publisher}{Springer Intern.}, \bibinfo{address}{Cham}, Chapter~13,
  \bibinfo{pages}{291--330}.
\newblock
\href{https://doi.org/10.1007/978-3-319-26633-6_13}{doi:\nolinkurl{10.1007/978-3-319-26633-6_13}}


\bibitem[Drepper(2007)]%
        {memory}
\bibfield{author}{\bibinfo{person}{Ulrich Drepper}.}
  \bibinfo{year}{2007}\natexlab{}.
\newblock \bibinfo{booktitle}{\emph{What every programmer should know about
  memory}}.
\newblock \bibinfo{type}{{T}echnical {R}eport}. \bibinfo{institution}{Red Hat,
  Inc}.
\newblock
\urldef\tempurl%
\url{https://people.freebsd.org/~lstewart/articles/cpumemory.pdf}
\showURL{%
\tempurl}


\bibitem[Du et~al\mbox{.}(2017)]%
        {Du:2017_TVCG}
\bibfield{author}{\bibinfo{person}{Fan Du}, \bibinfo{person}{Ben Shneiderman},
  \bibinfo{person}{Catherine Plaisant}, \bibinfo{person}{Sana Malik}, {and}
  \bibinfo{person}{Adam Perer}.} \bibinfo{year}{2017}\natexlab{}.
\newblock \showarticletitle{{Coping with Volume and Variety in Temporal Event
  Sequences: Strategies for Sharpening Analytic Focus}}.
\newblock \bibinfo{journal}{\emph{{IEEE} Trans. Vis. Comput. Graph.}}
  \bibinfo{volume}{23}, \bibinfo{number}{6} (\bibinfo{year}{2017}),
  \bibinfo{pages}{1636--1649}.
\newblock
\href{https://doi.org/10.1109/TVCG.2016.2539960}{doi:\nolinkurl{10.1109/TVCG.2016.2539960}}


\bibitem[Fekete et~al\mbox{.}(2024)]%
        {PDABook}
\bibfield{author}{\bibinfo{person}{Jean-Daniel Fekete}, \bibinfo{person}{Danyel
  Fisher}, {and} \bibinfo{person}{Michael Sedlmair}.}
  \bibinfo{year}{2024}\natexlab{}.
\newblock \bibinfo{booktitle}{\emph{Progressive Data Analysis: Roadmap and
  Research Agenda}}.
\newblock \bibinfo{publisher}{Eurographics}, \bibinfo{address}{Eindhoven, The
  Netherlands}. 231 pages.
\newblock
\showISBNx{978-3-03868-270-7}
\href{https://doi.org/10.2312/pda.20242707}{doi:\nolinkurl{10.2312/pda.20242707}}


\bibitem[Fekete and Freire(2020)]%
        {reproducibility}
\bibfield{author}{\bibinfo{person}{Jean-Daniel Fekete} {and}
  \bibinfo{person}{Juliana Freire}.} \bibinfo{year}{2020}\natexlab{}.
\newblock \showarticletitle{{Exploring Reproducibility in Visualization}}.
\newblock \bibinfo{journal}{\emph{{IEEE} Computer Graphics and Applications}}
  \bibinfo{volume}{40}, \bibinfo{number}{5} (\bibinfo{year}{2020}),
  \bibinfo{pages}{108--119}.
\newblock
\href{https://doi.org/10.1109/MCG.2020.3006412}{doi:\nolinkurl{10.1109/MCG.2020.3006412}}


\bibitem[Fisher et~al\mbox{.}(2012)]%
        {Fisher:2012_CGA}
\bibfield{author}{\bibinfo{person}{Danyel Fisher}, \bibinfo{person}{Steven~Mark
  Drucker}, {and} \bibinfo{person}{Arnd~Christian K{\"{o}}nig}.}
  \bibinfo{year}{2012}\natexlab{}.
\newblock \showarticletitle{Exploratory Visualization Involving Incremental,
  Approximate Database Queries and Uncertainty}.
\newblock \bibinfo{journal}{\emph{{IEEE} Computer Graphics and Applications}}
  \bibinfo{volume}{32}, \bibinfo{number}{4} (\bibinfo{year}{2012}),
  \bibinfo{pages}{55--62}.
\newblock
\href{https://doi.org/10.1109/MCG.2012.48}{doi:\nolinkurl{10.1109/MCG.2012.48}}


\bibitem[Fung et~al\mbox{.}(2010)]%
        {Fung:2010:CS}
\bibfield{author}{\bibinfo{person}{Benjamin C.~M. Fung}, \bibinfo{person}{Ke
  Wang}, \bibinfo{person}{Rui Chen}, {and} \bibinfo{person}{Philip~S. Yu}.}
  \bibinfo{year}{2010}\natexlab{}.
\newblock \showarticletitle{{Privacy-Preserving Data Publishing: A Survey of
  Recent Developments}}.
\newblock \bibinfo{journal}{\emph{ACM Comput. Surv.}} \bibinfo{volume}{42},
  \bibinfo{number}{4}, Article \bibinfo{articleno}{14} (\bibinfo{date}{June}
  \bibinfo{year}{2010}), \bibinfo{numpages}{53}~pages.
\newblock
\showISSN{0360-0300}
\href{https://doi.org/10.1145/1749603.1749605}{doi:\nolinkurl{10.1145/1749603.1749605}}


\bibitem[Goncalves et~al\mbox{.}(2020)]%
        {Goncalves:2020:BMC}
\bibfield{author}{\bibinfo{person}{Andre Goncalves}, \bibinfo{person}{Priyadip
  Ray}, \bibinfo{person}{Braden Soper}, \bibinfo{person}{Jennifer Stevens},
  \bibinfo{person}{Linda Coyle}, {and} \bibinfo{person}{Ana~Paula Sales}.}
  \bibinfo{year}{2020}\natexlab{}.
\newblock \showarticletitle{Generation and evaluation of synthetic patient
  data}.
\newblock \bibinfo{journal}{\emph{BMC Medical Research Methodology}}
  \bibinfo{volume}{20}, \bibinfo{number}{108} (\bibinfo{year}{2020}).
\newblock
\href{https://doi.org/10.1186/s12874-020-00977-1}{doi:\nolinkurl{10.1186/s12874-020-00977-1}}


\bibitem[Gotz and Stavropoulos(2014)]%
        {gotz2014decisionflow}
\bibfield{author}{\bibinfo{person}{David Gotz} {and} \bibinfo{person}{Harry
  Stavropoulos}.} \bibinfo{year}{2014}\natexlab{}.
\newblock \showarticletitle{DecisionFlow: Visual Analytics for High-Dimensional
  Temporal Event Sequence Data}.
\newblock \bibinfo{journal}{\emph{{IEEE} Trans. Vis. Comput. Graph.}}
  \bibinfo{volume}{20}, \bibinfo{number}{12} (\bibinfo{year}{2014}),
  \bibinfo{pages}{1783--1792}.
\newblock
\href{https://doi.org/10.1109/TVCG.2014.2346682}{doi:\nolinkurl{10.1109/TVCG.2014.2346682}}


\bibitem[Guan et~al\mbox{.}(2021)]%
        {Guan:2021:TCBB}
\bibfield{author}{\bibinfo{person}{Jiaqi Guan}, \bibinfo{person}{Runzhe Li},
  \bibinfo{person}{Sheng Yu}, {and} \bibinfo{person}{Xuegong Zhang}.}
  \bibinfo{year}{2021}\natexlab{}.
\newblock \showarticletitle{A Method for Generating Synthetic Electronic
  Medical Record Text}.
\newblock \bibinfo{journal}{\emph{IEEE/ACM Transactions on Computational
  Biology and Bioinformatics}} \bibinfo{volume}{18}, \bibinfo{number}{1}
  (\bibinfo{year}{2021}), \bibinfo{pages}{173--182}.
\newblock
\href{https://doi.org/10.1109/TCBB.2019.2948985}{doi:\nolinkurl{10.1109/TCBB.2019.2948985}}


\bibitem[Guo et~al\mbox{.}(2018)]%
        {GuoEventThread}
\bibfield{author}{\bibinfo{person}{Shunan Guo}, \bibinfo{person}{Ke Xu},
  \bibinfo{person}{Rongwen Zhao}, \bibinfo{person}{David Gotz},
  \bibinfo{person}{Hongyuan Zha}, {and} \bibinfo{person}{Nan Cao}.}
  \bibinfo{year}{2018}\natexlab{}.
\newblock \showarticletitle{EventThread: Visual Summarization and Stage
  Analysis of Event Sequence Data}.
\newblock \bibinfo{journal}{\emph{{IEEE} Trans. Vis. Comput. Graph.}}
  \bibinfo{volume}{24}, \bibinfo{number}{1} (\bibinfo{year}{2018}),
  \bibinfo{pages}{56--65}.
\newblock
\href{https://doi.org/10.1109/TVCG.2017.2745320}{doi:\nolinkurl{10.1109/TVCG.2017.2745320}}


\bibitem[Heer and Moritz(2024)]%
        {Mosaic}
\bibfield{author}{\bibinfo{person}{Jeffrey Heer} {and} \bibinfo{person}{Dominik
  Moritz}.} \bibinfo{year}{2024}\natexlab{}.
\newblock \showarticletitle{Mosaic: An Architecture for Scalable {\&}
  Interoperable Data Views}.
\newblock \bibinfo{journal}{\emph{{IEEE} Trans. Vis. Comput. Graph.}}
  \bibinfo{volume}{30}, \bibinfo{number}{1} (\bibinfo{year}{2024}),
  \bibinfo{pages}{436--446}.
\newblock
\href{https://doi.org/10.1109/TVCG.2023.3327189}{doi:\nolinkurl{10.1109/TVCG.2023.3327189}}


\bibitem[Jiang et~al\mbox{.}(2016)]%
        {jiang2016healthcare}
\bibfield{author}{\bibinfo{person}{Shenhui Jiang}, \bibinfo{person}{Shiaofen
  Fang}, \bibinfo{person}{Sam Bloomquist}, \bibinfo{person}{Jeremy Keiper},
  \bibinfo{person}{Mathew~J Palakal}, \bibinfo{person}{Yuni Xia}, {and}
  \bibinfo{person}{Shaun~J Grannis}.} \bibinfo{year}{2016}\natexlab{}.
\newblock \showarticletitle{Healthcare Data Visualization: Geospatial and
  Temporal Integration}. In \bibinfo{booktitle}{\emph{VISIGRAPP (2: IVAPP)}}.
  \bibinfo{publisher}{SCITEPRESS}, \bibinfo{address}{Setúbal, Portugal},
  \bibinfo{pages}{214--221}.
\newblock


\bibitem[Klemm et~al\mbox{.}(2015)]%
        {klemm20153d}
\bibfield{author}{\bibinfo{person}{Paul Klemm}, \bibinfo{person}{Kai Lawonn},
  \bibinfo{person}{Sylvia Gla{\ss}er}, \bibinfo{person}{Uli Niemann},
  \bibinfo{person}{Katrin Hegenscheid}, \bibinfo{person}{Henry V{\"o}lzke},
  {and} \bibinfo{person}{Bernhard Preim}.} \bibinfo{year}{2015}\natexlab{}.
\newblock \showarticletitle{3D regression heat map analysis of population study
  data}.
\newblock \bibinfo{journal}{\emph{{IEEE} Trans. Vis. Comput. Graph.}}
  \bibinfo{volume}{22}, \bibinfo{number}{1} (\bibinfo{year}{2015}),
  \bibinfo{pages}{81--90}.
\newblock


\bibitem[Kokosi and Harron(2022)]%
        {Kokosie:2022:BMJ}
\bibfield{author}{\bibinfo{person}{Theodora Kokosi} {and}
  \bibinfo{person}{Katie Harron}.} \bibinfo{year}{2022}\natexlab{}.
\newblock \showarticletitle{Synthetic data in medical research}.
\newblock \bibinfo{journal}{\emph{BMJ Medicine}} \bibinfo{volume}{1},
  \bibinfo{number}{1} (\bibinfo{year}{2022}).
\newblock
\href{https://doi.org/10.1136/bmjmed-2022-000167}{doi:\nolinkurl{10.1136/bmjmed-2022-000167}}


\bibitem[Li et~al\mbox{.}(2007)]%
        {Li:2007:ICDE}
\bibfield{author}{\bibinfo{person}{Ninghui Li}, \bibinfo{person}{Tiancheng Li},
  {and} \bibinfo{person}{Suresh Venkatasubramanian}.}
  \bibinfo{year}{2007}\natexlab{}.
\newblock \showarticletitle{{t-Closeness: Privacy Beyond k-Anonymity and
  l-Diversity}}. In \bibinfo{booktitle}{\emph{Proc.\ ICDE}}.
  \bibinfo{publisher}{IEEE}, \bibinfo{address}{Los Alamitos},
  \bibinfo{pages}{106--115}.
\newblock
\href{https://doi.org/10.1109/ICDE.2007.367856}{doi:\nolinkurl{10.1109/ICDE.2007.367856}}


\bibitem[Liu and Heer(2014)]%
        {Liu:2014_TVCG}
\bibfield{author}{\bibinfo{person}{Zhicheng Liu} {and} \bibinfo{person}{Jeffrey
  Heer}.} \bibinfo{year}{2014}\natexlab{}.
\newblock \showarticletitle{{The Effects of Interactive Latency on Exploratory
  Visual Analysis}}.
\newblock \bibinfo{journal}{\emph{{IEEE} Trans. Vis. Comput. Graph.}}
  \bibinfo{volume}{20}, \bibinfo{number}{12} (\bibinfo{year}{2014}),
  \bibinfo{pages}{2122--2131}.
\newblock
\href{https://doi.org/10.1109/TVCG.2014.2346452}{doi:\nolinkurl{10.1109/TVCG.2014.2346452}}


\bibitem[Liu et~al\mbox{.}(2013)]%
        {imMens}
\bibfield{author}{\bibinfo{person}{Zhicheng Liu}, \bibinfo{person}{Biye Jiang},
  {and} \bibinfo{person}{Jeffrey Heer}.} \bibinfo{year}{2013}\natexlab{}.
\newblock \showarticletitle{\emph{imMens}: Real-time Visual Querying of Big
  Data}.
\newblock \bibinfo{journal}{\emph{Comput. Graph. Forum}} \bibinfo{volume}{32},
  \bibinfo{number}{3} (\bibinfo{year}{2013}), \bibinfo{pages}{421--430}.
\newblock
\href{https://doi.org/10.1111/CGF.12129}{doi:\nolinkurl{10.1111/CGF.12129}}


\bibitem[Liu et~al\mbox{.}(2016)]%
        {liu2016patterns}
\bibfield{author}{\bibinfo{person}{Zhicheng Liu}, \bibinfo{person}{Yang Wang},
  \bibinfo{person}{Mira Dontcheva}, \bibinfo{person}{Matthew Hoffman},
  \bibinfo{person}{Seth Walker}, {and} \bibinfo{person}{Alan Wilson}.}
  \bibinfo{year}{2016}\natexlab{}.
\newblock \showarticletitle{Patterns and sequences: Interactive exploration of
  clickstreams to understand common visitor paths}.
\newblock \bibinfo{journal}{\emph{{IEEE} Trans. Vis. Comput. Graph.}}
  \bibinfo{volume}{23}, \bibinfo{number}{1} (\bibinfo{year}{2016}),
  \bibinfo{pages}{321--330}.
\newblock


\bibitem[Liu et~al\mbox{.}(2017)]%
        {DBLP_journals/tvcg/LiuWDHWW17}
\bibfield{author}{\bibinfo{person}{Zhicheng Liu}, \bibinfo{person}{Yang Wang},
  \bibinfo{person}{Mira Dontcheva}, \bibinfo{person}{Matthew Hoffman},
  \bibinfo{person}{Seth Walker}, {and} \bibinfo{person}{Alan Wilson}.}
  \bibinfo{year}{2017}\natexlab{}.
\newblock \showarticletitle{Patterns and Sequences: Interactive Exploration of
  Clickstreams to Understand Common Visitor Paths}.
\newblock \bibinfo{journal}{\emph{{IEEE} Trans. Vis. Comput. Graph.}}
  \bibinfo{volume}{23}, \bibinfo{number}{1} (\bibinfo{year}{2017}),
  \bibinfo{pages}{321--330}.
\newblock
\href{https://doi.org/10.1109/TVCG.2016.2598797}{doi:\nolinkurl{10.1109/TVCG.2016.2598797}}


\bibitem[Meyer et~al\mbox{.}(2013)]%
        {Meyer:2013_VAHC}
\bibfield{author}{\bibinfo{person}{Tamra~E. Meyer}, \bibinfo{person}{Megan
  Monroe}, \bibinfo{person}{Catherine Plaisant}, \bibinfo{person}{Rongjian
  Lan}, \bibinfo{person}{Krist Wongsuphasawat}, \bibinfo{person}{Trinka~S.
  Coster}, \bibinfo{person}{Sigfried Gold}, \bibinfo{person}{Jeff Millstein},
  {and} \bibinfo{person}{Ben Shneiderman}.} \bibinfo{year}{2013}\natexlab{}.
\newblock \showarticletitle{{Visualizing Patterns of Drug Prescriptions with
  {EventFlow}: A Pilot Study of Asthma Medications in the Military Health
  System}}. In \bibinfo{booktitle}{\emph{Proc.\ VAHC}}.
  \bibinfo{publisher}{ACM}, \bibinfo{address}{New York},
  \bibinfo{pages}{55--58}.
\newblock
\urldef\tempurl%
\url{https://www.visualanalyticshealthcare.org/docs/VAHC2013_proceedings.pdf}
\showURL{%
\tempurl}


\bibitem[Monroe et~al\mbox{.}(2013)]%
        {Monroe:2013_TVCG}
\bibfield{author}{\bibinfo{person}{Megan Monroe}, \bibinfo{person}{Rongjian
  Lan}, \bibinfo{person}{Hanseung Lee}, \bibinfo{person}{Catherine Plaisant},
  {and} \bibinfo{person}{Ben Shneiderman}.} \bibinfo{year}{2013}\natexlab{}.
\newblock \showarticletitle{{Temporal Event Sequence Simplification}}.
\newblock \bibinfo{journal}{\emph{{IEEE} Trans. Vis. Comput. Graph.}}
  \bibinfo{volume}{19}, \bibinfo{number}{12} (\bibinfo{year}{2013}),
  \bibinfo{pages}{2227--2236}.
\newblock
\href{https://doi.org/10.1109/TVCG.2013.200}{doi:\nolinkurl{10.1109/TVCG.2013.200}}


\bibitem[Nielsen(1993)]%
        {Nielsen:1993_Book}
\bibfield{author}{\bibinfo{person}{Jakob Nielsen}.}
  \bibinfo{year}{1993}\natexlab{}.
\newblock \showarticletitle{{Usability Heuristics}}.
\newblock In \bibinfo{booktitle}{\emph{Usability Engineering}}.
  \bibinfo{publisher}{Morgan Kaufmann}, \bibinfo{address}{San Diego},
  Chapter~5, \bibinfo{pages}{115--163}.
\newblock
\showISBNx{978-0-12-518406-9}
\href{https://doi.org/10.1016/B978-0-08-052029-2.50008-5}{doi:\nolinkurl{10.1016/B978-0-08-052029-2.50008-5}}


\bibitem[Ozkaynak et~al\mbox{.}(2015)]%
        {Ozkaynak:2015_BI}
\bibfield{author}{\bibinfo{person}{Mustafa Ozkaynak}, \bibinfo{person}{Oliwier
  Dziadkowiec}, \bibinfo{person}{Rakesh Mistry}, \bibinfo{person}{Tiffany
  Callahan}, \bibinfo{person}{Ze He}, \bibinfo{person}{Sara Deakyne}, {and}
  \bibinfo{person}{Eric Tham}.} \bibinfo{year}{2015}\natexlab{}.
\newblock \showarticletitle{Characterizing workflow for pediatric asthma
  patients in emergency departments using electronic health records}.
\newblock \bibinfo{journal}{\emph{Journal of Biomedical Informatics}}
  \bibinfo{volume}{57} (\bibinfo{year}{2015}), \bibinfo{pages}{386--398}.
\newblock
\showISSN{1532-0464}
\href{https://doi.org/10.1016/j.jbi.2015.08.018}{doi:\nolinkurl{10.1016/j.jbi.2015.08.018}}


\bibitem[Patil et~al\mbox{.}(2023)]%
        {Patil:2023_TVCG}
\bibfield{author}{\bibinfo{person}{Ameya Patil}, \bibinfo{person}{Gaëlle
  Richer}, \bibinfo{person}{Christopher Jermaine}, \bibinfo{person}{Dominik
  Moritz}, {and} \bibinfo{person}{Jean-Daniel Fekete}.}
  \bibinfo{year}{2023}\natexlab{}.
\newblock \showarticletitle{{Studying Early Decision Making with Progressive
  Bar Charts}}.
\newblock \bibinfo{journal}{\emph{{IEEE} Trans. Vis. Comput. Graph.}}
  \bibinfo{volume}{29}, \bibinfo{number}{1} (\bibinfo{year}{2023}),
  \bibinfo{pages}{407--417}.
\newblock
\href{https://doi.org/10.1109/TVCG.2022.3209426}{doi:\nolinkurl{10.1109/TVCG.2022.3209426}}


\bibitem[Perer and Wang(2014a)]%
        {DBLP_conf/iui/PererW14}
\bibfield{author}{\bibinfo{person}{Adam Perer} {and} \bibinfo{person}{Fei
  Wang}.} \bibinfo{year}{2014}\natexlab{a}.
\newblock \showarticletitle{Frequence: interactive mining and visualization of
  temporal frequent event sequences}. In \bibinfo{booktitle}{\emph{19th
  International Conference on Intelligent User Interfaces, {IUI} 2014, Haifa,
  Israel, February 24-27, 2014}}, \bibfield{editor}{\bibinfo{person}{Tsvi
  Kuflik}, \bibinfo{person}{Oliviero Stock}, \bibinfo{person}{Joyce~Yue Chai},
  {and} \bibinfo{person}{Antonio Kr{\"{u}}ger}} (Eds.).
  \bibinfo{publisher}{{ACM}}, \bibinfo{address}{New York},
  \bibinfo{pages}{153--162}.
\newblock
\href{https://doi.org/10.1145/2557500.2557508}{doi:\nolinkurl{10.1145/2557500.2557508}}


\bibitem[Perer and Wang(2014b)]%
        {perer2014frequence}
\bibfield{author}{\bibinfo{person}{Adam Perer} {and} \bibinfo{person}{Fei
  Wang}.} \bibinfo{year}{2014}\natexlab{b}.
\newblock \showarticletitle{Frequence: interactive mining and visualization of
  temporal frequent event sequences}. In \bibinfo{booktitle}{\emph{19th
  International Conference on Intelligent User Interfaces, {IUI} 2014, Haifa,
  Israel, February 24-27, 2014}}, \bibfield{editor}{\bibinfo{person}{Tsvi
  Kuflik}, \bibinfo{person}{Oliviero Stock}, \bibinfo{person}{Joyce~Yue Chai},
  {and} \bibinfo{person}{Antonio Kr{\"{u}}ger}} (Eds.).
  \bibinfo{publisher}{{ACM}}, \bibinfo{address}{New York},
  \bibinfo{pages}{153--162}.
\newblock
\href{https://doi.org/10.1145/2557500.2557508}{doi:\nolinkurl{10.1145/2557500.2557508}}


\bibitem[Ping et~al\mbox{.}(2017)]%
        {Ping:2017:SSDBM}
\bibfield{author}{\bibinfo{person}{Haoyue Ping}, \bibinfo{person}{Julia
  Stoyanovich}, {and} \bibinfo{person}{Bill Howe}.}
  \bibinfo{year}{2017}\natexlab{}.
\newblock \showarticletitle{{DataSynthesizer}: Privacy-Preserving Synthetic
  Datasets}. In \bibinfo{booktitle}{\emph{Proc.\ SSDBM}}.
  \bibinfo{publisher}{ACM}, \bibinfo{address}{New York},
  \bibinfo{pages}{1:42--5:42}.
\newblock
\showISBNx{9781450352826}
\href{https://doi.org/10.1145/3085504.3091117}{doi:\nolinkurl{10.1145/3085504.3091117}}


\bibitem[Pister et~al\mbox{.}(2023)]%
        {Pister:2023_CGF}
\bibfield{author}{\bibinfo{person}{Alexis Pister}, \bibinfo{person}{Christophe
  Prieur}, {and} \bibinfo{person}{Jean{-}Daniel Fekete}.}
  \bibinfo{year}{2023}\natexlab{}.
\newblock \showarticletitle{ComBiNet: Visual Query and Comparison of Bipartite
  Multivariate Dynamic Social Networks}.
\newblock \bibinfo{journal}{\emph{Comput. Graph. Forum}} \bibinfo{volume}{42},
  \bibinfo{number}{1} (\bibinfo{year}{2023}), \bibinfo{pages}{290--304}.
\newblock
\href{https://doi.org/10.1111/CGF.14731}{doi:\nolinkurl{10.1111/CGF.14731}}


\bibitem[Raasveldt and M{\"{u}}hleisen(2019)]%
        {DuckDB}
\bibfield{author}{\bibinfo{person}{Mark Raasveldt} {and}
  \bibinfo{person}{Hannes M{\"{u}}hleisen}.} \bibinfo{year}{2019}\natexlab{}.
\newblock \showarticletitle{DuckDB: an Embeddable Analytical Database}. In
  \bibinfo{booktitle}{\emph{Proceedings of the 2019 International Conference on
  Management of Data, {SIGMOD} Conference}},
  \bibfield{editor}{\bibinfo{person}{Peter~A. Boncz}, \bibinfo{person}{Stefan
  Manegold}, \bibinfo{person}{Anastasia Ailamaki}, \bibinfo{person}{Amol
  Deshpande}, {and} \bibinfo{person}{Tim Kraska}} (Eds.).
  \bibinfo{publisher}{{ACM}}, \bibinfo{address}{New York},
  \bibinfo{pages}{1981--1984}.
\newblock
\href{https://doi.org/10.1145/3299869.3320212}{doi:\nolinkurl{10.1145/3299869.3320212}}


\bibitem[Richer et~al\mbox{.}(2024a)]%
        {PDABook:9}
\bibfield{author}{\bibinfo{person}{Ga\"elle Richer},
  \bibinfo{person}{Jean-Daniel Fekete}, {and} \bibinfo{person}{Michael
  Sedlmair}.} \bibinfo{year}{2024}\natexlab{a}.
\newblock \showarticletitle{Evaluation for Progressive Data Analysis}.
\newblock In \bibinfo{booktitle}{\emph{Progressive Data Analysis: Roadmap and
  Research Agenda}}, \bibfield{editor}{\bibinfo{person}{Jean-Daniel Fekete},
  \bibinfo{person}{Danyel Fisher}, {and} \bibinfo{person}{Michael Sedlmair}}
  (Eds.). \bibinfo{publisher}{Eurographics}, \bibinfo{address}{Eindhoven, The
  Netherlands}, \bibinfo{pages}{149--170}.
\newblock
\showISBNx{978-3-03868-270-7}
\href{https://doi.org/10.2312/pda.20242707}{doi:\nolinkurl{10.2312/pda.20242707}}


\bibitem[Richer et~al\mbox{.}(2024b)]%
        {Richer:2022_TVCG}
\bibfield{author}{\bibinfo{person}{Ga{\"{e}}lle Richer},
  \bibinfo{person}{Alexis Pister}, \bibinfo{person}{Moataz Abdelaal},
  \bibinfo{person}{Jean{-}Daniel Fekete}, \bibinfo{person}{Michael Sedlmair},
  {and} \bibinfo{person}{Daniel Weiskopf}.} \bibinfo{year}{2024}\natexlab{b}.
\newblock \showarticletitle{Scalability in Visualization}.
\newblock \bibinfo{journal}{\emph{{IEEE} Trans. Vis. Comput. Graph.}}
  \bibinfo{volume}{30}, \bibinfo{number}{7} (\bibinfo{year}{2024}),
  \bibinfo{pages}{3314--3330}.
\newblock
\href{https://doi.org/10.1109/TVCG.2022.3231230}{doi:\nolinkurl{10.1109/TVCG.2022.3231230}}


\bibitem[Rusu et~al\mbox{.}(2024)]%
        {PDABook:3}
\bibfield{author}{\bibinfo{person}{Florin Rusu}, \bibinfo{person}{Carsten
  Binnig}, {and} \bibinfo{person}{Chris Weaver}.}
  \bibinfo{year}{2024}\natexlab{}.
\newblock \showarticletitle{Data Management for Progressive Data Analysis}.
\newblock In \bibinfo{booktitle}{\emph{Progressive Data Analysis: Roadmap and
  Research Agenda}}, \bibfield{editor}{\bibinfo{person}{Jean-Daniel Fekete},
  \bibinfo{person}{Danyel Fisher}, {and} \bibinfo{person}{Michael Sedlmair}}
  (Eds.). \bibinfo{publisher}{Eurographics}, \bibinfo{address}{Eindhoven, The
  Netherlands}, \bibinfo{pages}{33--48}.
\newblock
\showISBNx{978-3-03868-270-7}
\href{https://doi.org/10.2312/pda.20242707}{doi:\nolinkurl{10.2312/pda.20242707}}


\bibitem[Shneiderman(1996)]%
        {Mantra}
\bibfield{author}{\bibinfo{person}{Ben Shneiderman}.}
  \bibinfo{year}{1996}\natexlab{}.
\newblock \showarticletitle{The eyes have it: a task by data type taxonomy for
  information visualizations}. In \bibinfo{booktitle}{\emph{Proceedings 1996
  IEEE Symposium on Visual Languages}}. \bibinfo{publisher}{IEEE},
  \bibinfo{address}{Piscataway, NJ}, \bibinfo{pages}{336--343}.
\newblock
\href{https://doi.org/10.1109/VL.1996.545307}{doi:\nolinkurl{10.1109/VL.1996.545307}}


\bibitem[SNDS(2023)]%
        {SNDS}
SNDS \bibinfo{year}{2023}\natexlab{}.
\newblock \bibinfo{title}{{Système National des Données de Santé (SNDS)}}.
\newblock \bibinfo{howpublished}{\url{https://www.snds.gouv.fr/}}.
\newblock
\newblock
\shownote{Accessed: 2023-02-01}.


\bibitem[Stolper et~al\mbox{.}(2014)]%
        {Stolper:2014_TVCG}
\bibfield{author}{\bibinfo{person}{Charles~D. Stolper}, \bibinfo{person}{Adam
  Perer}, {and} \bibinfo{person}{David Gotz}.} \bibinfo{year}{2014}\natexlab{}.
\newblock \showarticletitle{{Progressive Visual Analytics: User-Driven Visual
  Exploration of In-Progress Analytics}}.
\newblock \bibinfo{journal}{\emph{{IEEE} Trans. Vis. Comput. Graph.}}
  \bibinfo{volume}{20}, \bibinfo{number}{12} (\bibinfo{year}{2014}),
  \bibinfo{pages}{1653--1662}.
\newblock
\href{https://doi.org/10.1109/TVCG.2014.2346574}{doi:\nolinkurl{10.1109/TVCG.2014.2346574}}


\bibitem[UCI ML({[n.\,d.]})]%
        {uci-ml}
UCI ML \bibinfo{year}{[n.\,d.]}\natexlab{}.
\newblock \bibinfo{title}{The UCI Machine Learning Repository}.
\newblock \bibinfo{howpublished}{\url{https://archive.ics.uci.edu}}.
\newblock
\newblock
\shownote{Accessed: 2025-02-23}.


\bibitem[Ulmer et~al\mbox{.}(2024)]%
        {Ulmer:2024_TVCG}
\bibfield{author}{\bibinfo{person}{Alex Ulmer}, \bibinfo{person}{Marco
  Angelini}, \bibinfo{person}{Jean-Daniel Fekete}, \bibinfo{person}{Jörn
  Kohlhammer}, {and} \bibinfo{person}{Thorsten May}.}
  \bibinfo{year}{2024}\natexlab{}.
\newblock \showarticletitle{{A Survey on Progressive Visualization}}.
\newblock \bibinfo{journal}{\emph{{IEEE} Trans. Vis. Comput. Graph.}}
  \bibinfo{volume}{30}, \bibinfo{number}{9} (\bibinfo{year}{2024}),
  \bibinfo{pages}{6447--6467}.
\newblock
\href{https://doi.org/10.1109/TVCG.2023.3346641}{doi:\nolinkurl{10.1109/TVCG.2023.3346641}}


\bibitem[Vilanova et~al\mbox{.}(2024)]%
        {PDABook:6}
\bibfield{author}{\bibinfo{person}{Anna Vilanova}, \bibinfo{person}{Marco
  Angelini}, \bibinfo{person}{Sriram~Karthik Badam}, {and}
  \bibinfo{person}{Jean-Daniel Fekete}.} \bibinfo{year}{2024}\natexlab{}.
\newblock \showarticletitle{Uncertainty and Quality for Progressive Data
  Analysis}.
\newblock In \bibinfo{booktitle}{\emph{Progressive Data Analysis: Roadmap and
  Research Agenda}}, \bibfield{editor}{\bibinfo{person}{Jean-Daniel Fekete},
  \bibinfo{person}{Danyel Fisher}, {and} \bibinfo{person}{Michael Sedlmair}}
  (Eds.). \bibinfo{publisher}{Eurographics}, \bibinfo{address}{Eindhoven, The
  Netherlands}, \bibinfo{pages}{92--107}.
\newblock
\showISBNx{978-3-03868-270-7}
\href{https://doi.org/10.2312/pda.20242707}{doi:\nolinkurl{10.2312/pda.20242707}}


\bibitem[Vrotsou and Nordman(2019)]%
        {DBLP_journals/tvcg/VrotsouN19}
\bibfield{author}{\bibinfo{person}{Katerina Vrotsou} {and}
  \bibinfo{person}{Aida Nordman}.} \bibinfo{year}{2019}\natexlab{}.
\newblock \showarticletitle{Exploratory Visual Sequence Mining Based on
  Pattern-Growth}.
\newblock \bibinfo{journal}{\emph{{IEEE} Trans. Vis. Comput. Graph.}}
  \bibinfo{volume}{25}, \bibinfo{number}{8} (\bibinfo{year}{2019}),
  \bibinfo{pages}{2597--2610}.
\newblock
\href{https://doi.org/10.1109/TVCG.2018.2848247}{doi:\nolinkurl{10.1109/TVCG.2018.2848247}}


\bibitem[Vue(2023)]%
        {Vue}
Vue \bibinfo{year}{2023}\natexlab{}.
\newblock \bibinfo{title}{{Vue: The Progressive JavaScript Framework}}.
\newblock \bibinfo{howpublished}{\url{https://vuejs.org/}}.
\newblock
\newblock
\shownote{Accessed: 2023-03-21}.


\bibitem[Walonoski et~al\mbox{.}(2020)]%
        {Walonoski:2020:IBM}
\bibfield{author}{\bibinfo{person}{Jason Walonoski}, \bibinfo{person}{Sybil
  Klaus}, \bibinfo{person}{Eldesia Granger}, \bibinfo{person}{Dylan Hall},
  \bibinfo{person}{Andrew Gregorowicz}, \bibinfo{person}{George Neyarapally},
  \bibinfo{person}{Abigail Watson}, {and} \bibinfo{person}{Jeff Eastman}.}
  \bibinfo{year}{2020}\natexlab{}.
\newblock \showarticletitle{Synthea\texttrademark\xspace Novel coronavirus
  ({COVID-19}) model and synthetic data set}.
\newblock \bibinfo{journal}{\emph{Intelligence-Based Medicine}}
  \bibinfo{volume}{1-2}, \bibinfo{number}{100007} (\bibinfo{year}{2020}).
\newblock
\showISSN{2666-5212}
\href{https://doi.org/10.1016/j.ibmed.2020.100007}{doi:\nolinkurl{10.1016/j.ibmed.2020.100007}}


\bibitem[Wang and Laramee(2022)]%
        {Wang:2022_CGF}
\bibfield{author}{\bibinfo{person}{Qi-Ru Wang} {and} \bibinfo{person}{Robert~S.
  Laramee}.} \bibinfo{year}{2022}\natexlab{}.
\newblock \showarticletitle{{EHR STAR: The State-Of-the-Art in Interactive
  {EHR} Visualization}}.
\newblock \bibinfo{journal}{\emph{Comput. Graph. Forum}} \bibinfo{volume}{41},
  \bibinfo{number}{1} (\bibinfo{year}{2022}), \bibinfo{pages}{69--105}.
\newblock
\href{https://doi.org/10.1111/cgf.14424}{doi:\nolinkurl{10.1111/cgf.14424}}


\bibitem[Wongsuphasawat and Gotz(2012)]%
        {Wongsuphasawat:2012_TVCG}
\bibfield{author}{\bibinfo{person}{Krist Wongsuphasawat} {and}
  \bibinfo{person}{David Gotz}.} \bibinfo{year}{2012}\natexlab{}.
\newblock \showarticletitle{{Exploring Flow, Factors, and Outcomes of Temporal
  Event Sequences with the Outflow Visualization}}.
\newblock \bibinfo{journal}{\emph{{IEEE} Trans. Vis. Comput. Graph.}}
  \bibinfo{volume}{18}, \bibinfo{number}{12} (\bibinfo{year}{2012}),
  \bibinfo{pages}{2659--2668}.
\newblock
\href{https://doi.org/10.1109/TVCG.2012.225}{doi:\nolinkurl{10.1109/TVCG.2012.225}}


\bibitem[Wongsuphasawat et~al\mbox{.}(2011)]%
        {Wongsuphasawat:2011_CHI}
\bibfield{author}{\bibinfo{person}{Krist Wongsuphasawat},
  \bibinfo{person}{John~Alexis Guerra~G\'{o}mez}, \bibinfo{person}{Catherine
  Plaisant}, \bibinfo{person}{Taowei~David Wang}, \bibinfo{person}{Meirav
  Taieb-Maimon}, {and} \bibinfo{person}{Ben Shneiderman}.}
  \bibinfo{year}{2011}\natexlab{}.
\newblock \showarticletitle{{LifeFlow: Visualizing an Overview of Event
  Sequences}}. In \bibinfo{booktitle}{\emph{Proc.\ CHI}}.
  \bibinfo{publisher}{ACM}, \bibinfo{address}{New York},
  \bibinfo{pages}{1747–--1756}.
\newblock
\showISBNx{9781450302289}
\href{https://doi.org/10.1145/1978942.1979196}{doi:\nolinkurl{10.1145/1978942.1979196}}


\bibitem[Zgraggen et~al\mbox{.}(2017)]%
        {Zgraggen:2016_TVCG}
\bibfield{author}{\bibinfo{person}{Emanuel Zgraggen}, \bibinfo{person}{Alex
  Galakatos}, \bibinfo{person}{Andrew Crotty}, \bibinfo{person}{Jean-Daniel
  Fekete}, {and} \bibinfo{person}{Tim Kraska}.}
  \bibinfo{year}{2017}\natexlab{}.
\newblock \showarticletitle{{How Progressive Visualizations Affect Exploratory
  Analysis}}.
\newblock \bibinfo{journal}{\emph{{IEEE} Trans. Vis. Comput. Graph.}}
  \bibinfo{volume}{23}, \bibinfo{number}{8} (\bibinfo{year}{2017}),
  \bibinfo{pages}{1977--1987}.
\newblock
\href{https://doi.org/10.1109/TVCG.2016.2607714}{doi:\nolinkurl{10.1109/TVCG.2016.2607714}}


\end{thebibliography}

\appendix 

\onecolumn
\newpage 
 


\begin{jdfenv}
\section{Comparison with EventFlow}\label{sec_compareEventFlow}
 
We tried the EventFlow program with a subset of our dataset containing 10\,k, 20\,k, and 50\,k synthetic patients. In comparison, ParcoursVis can be used with tens of millions of patients without noticeable latency; the online demo \anonurl{https://parcoursvis.lisn.upsaclay.fr/} uses 10\,M patients in a web setting, and we use ParcoursVis with 20\,M patients. ParcoursVis supports datasets 100--1,000 times larger.
To provide a baseline to better understand the scalability of ParcoursVis, we compare it to EventFlow version v2.3.4 for three important tasks and features on the three dataset sizes. We run the java virtual machine with 40\,Gb of memory (\texttt{java -Xmx40G -jar EventFlow\_v2.3.4.jar}).

\begin{center}
\begin{tabular}{rrrrr}
\hline
Operation &      10\,k & 20\,k & 50\,k\\
\hline
Load &       \textbf{4\,s} &  12\,s & 40\,s \\
Visualize & 20\,s &  85\,s & 410\,s \\
Hide Event & \textbf{4\,s} &  16\,s & 96\,s \\
Show Event & 20\,s &  77\,s & 500\,s \\
Merge Event & 25\,s &  80\,s & 454\,s \\
Memory & 13\,G & 26\,G  & 28\,G  \\
\hline
\end{tabular}
\end{center}

\begin{itemize}
    \item Loading time: EventFlow can download a CSV file in a reasonable time for 10\,k and 20\,k patients, not more.
    \item Visualization time: Once loaded, the datasets take quite a long time to appear on screen; the aggregation and layout times are longer than the 10\,s limit.
    \item Hiding and showing back an event are not symmetrical. Hiding takes a relatively short time under 20\,k patients. Showing back the event takes a much longer time, always longer than the 10\,s limit.
    \item Interaction time: hovering the \IT is always instantaneous in EventFlow, but above 20k patients, any action that changes the prefix tree takes from 16 seconds to 10 minutes. 
    \item Features: EventFlow lets users interactively specify several parameters and aggregation rules. We measured merging/aggregating consecutive ``alphabloc'' events.
    ParcoursVis performs this aggregation in its main loop. It always takes more than 10\,s with EventFlow.
\end{itemize}

We also report the memory usage when running the Java virtual machine. Our measures show that EventFlow cannot be considered interactive above 10\,k patients, and even then, many operations take longer than 10\,s to complete. In contrast, ParcoursVis can perform all its operations with a latency of under a couple of seconds, even for 20\,M patients.

\end{jdfenv}

\section{Synthetic Data Generation}\label{sec_syntheticData}
All the figures in our article rely on synthetic data. As health records are sensitive by nature, data regulations forbid sharing them outside private and secure platforms, and only to authorized persons. If we want to share our project openly and allow the reproduction of our evaluations, we need to create a dataset statistically realistic while protecting the patients' anonymity.

We use our aggregated tree as a model to produce synthetic patients since it holds a simplified but accurate statistical profile of all our patients. We can indeed replay the probabilities stored in the tree to generate new patients from which, if we aggregate those synthetic patients again, we should get another statistically similar prefix tree.

\begin{lstfloat}[b]
  \begin{lstlisting}[language=C++,captionpos=b,caption={Structure of a tree Node},label={lst:node}]
struct Node {
  /* Data needed for ParcoursVis*/
  unsigned    count; // Number of sequences passing through this node
  EventType   name; // event name, such as phyto or surgery
  list<Node*> children; // children nodes
  Histogram duration_distribution; // duration distribution

  /* Data related to our medical use case. Not useful in our evaluation */
  Histogram              age_distribution; // age distribution
  map<Disease, unsigned> disease_count; // distribution of comorbidities
};
\end{lstlisting}
\end{lstfloat}

The tree contains nodes (\autoref{lst:node}), each node $N_i$ representing a pathway $P_i$ from the root to $N_i$:
\begin{itemize}[nosep]
    \item count: The number of patients who started their treatment with prefix $P_i$, reaching $N_i$.
    \item children: The list of events following $P_i$. For each event $E_j$, we export its frequency, \ie probability of $P({E_j}|{P_i})$.
    \item age\_distribution: The age distribution of the patients reaching $N_i$.
    \item The duration distribution of the last event (treatment) reaching $N_i$.
    \item The comorbidity distribution for the patients reaching $N_i$.
\end{itemize}

From this tree, we can generate a new patient in three phases: (1) generating a random sequence of high-level events from the tree, (2) adding realistic attributes to the high-level events, and (3) generating a plausible low-level event sequence from the high-level events. While it is common to rely on statistics and machine learning models~\cite{Goncalves:2020:BMC,  Walonoski:2020:IBM, Ping:2017:SSDBM, Guan:2021:TCBB, Kokosie:2022:BMJ} to generate synthetic datasets, to the best of our knowledge, we are the first ones to rely on a prefix tree to generate synthetic datasets.

\noindent\textit{\textbf{Random sequence: }} We start at the root node that becomes the current node. Iteratively, we chose the next event from the current node with a probability weighted by the event frequencies (\eg 40\% of chance to pick ``alphabloc'' in the root node of \autoref{fig_mainView_a}). This next event is appended to our random sequence. We iterate until the next event selected is ``end of treatment.''
At the end of this phase, we have a list of high-level events, and we also keep the list of the current nodes.

\noindent\textit{\textbf{Adding Attributes: }} The nodes created at line 24 of \autoref{lst_aggregation} by \texttt{getHighLevelEventSequence}  contain a name, an age, a duration, and possible comorbidities. We generate the same attributes for our events at this stage. We generate the attributes backward, starting from the last event associated with the last node that we kept previously. Using the age distribution of the last node, we generate a random age that we set to the last created event. Same for the duration, and from the frequency of comorbidities, we generate a comorbidity pattern. Then, we proceed backward to the event list and tree nodes to create our events, updating the age according to the generated duration, propagating the comorbidity pattern (we do not change it yet), and picking a random duration again. This process is repeated until we reach the first event. Our high-level sequence is then complete.

\noindent\textit{\textbf{Generating Low-Level Events: }} We first pick a plausible starting date at random for our low-level sequence. From the generated high-level sequence, we split events with duration into a sequence of atomic events (\eg drug purchases) at regular intervals (15 days for each box of pills). Each low-level event is then given a date. We also remove the synthetic ``interruption'' and ``no treatment'' events, updating the date of the next concrete event accordingly. This phase generates a plausible sequence of events for one synthetic patient, similar to the CSV file \autoref{tab_csv}.
  
\smallskip
Our strategy uses the statistical profile of each node; when we generate a large number of synthetic patients and aggregate them, the final tree is very close statistically to the original one. However, our approach does not preserve higher-level statistics, such as correlations or dependencies between different node statistics.
Also, we do not model the evolution of comorbidities; for each patient, we randomly select a set according to the last event and keep it during the whole sequence. More work is needed to take into account the appearance and progression of the comorbidities of aging patients. 

In the end, our synthetic data allows us to showcase our system, but we do not claim to model the patients accurately enough for other purposes.

\noindent\textit{\textbf{Anonymization}}
Generating synthetic patients seems to ensure patients' anonymity, but generating a large number of patients may disclose statistical information from the aggregated tree that is too revealing, as an aggregated tree from millions of synthetic patients can recreate small nodes.

To be able to share synthetic data, according to national regulations, we should process the aggregated tree. The regulations are different for different countries. 
As fully anonymizing the aggregated tree is out of the scope of this article, we instead give readers interested in that topic a few references that discuss the potential threats of attackers, and the solutions to minimize them with the cost of lesser diversity in the aggregated data~\cite{Li:2007:ICDE, Fung:2010:CS, Brickell:2008:KDD}.

%
%
%


\section{Additional Figures}

Here, we convey additional Figures related to \autoref{sec_eval}.

\begin{figure}[H]\centering
    \includegraphics[width=0.5\linewidth]
    {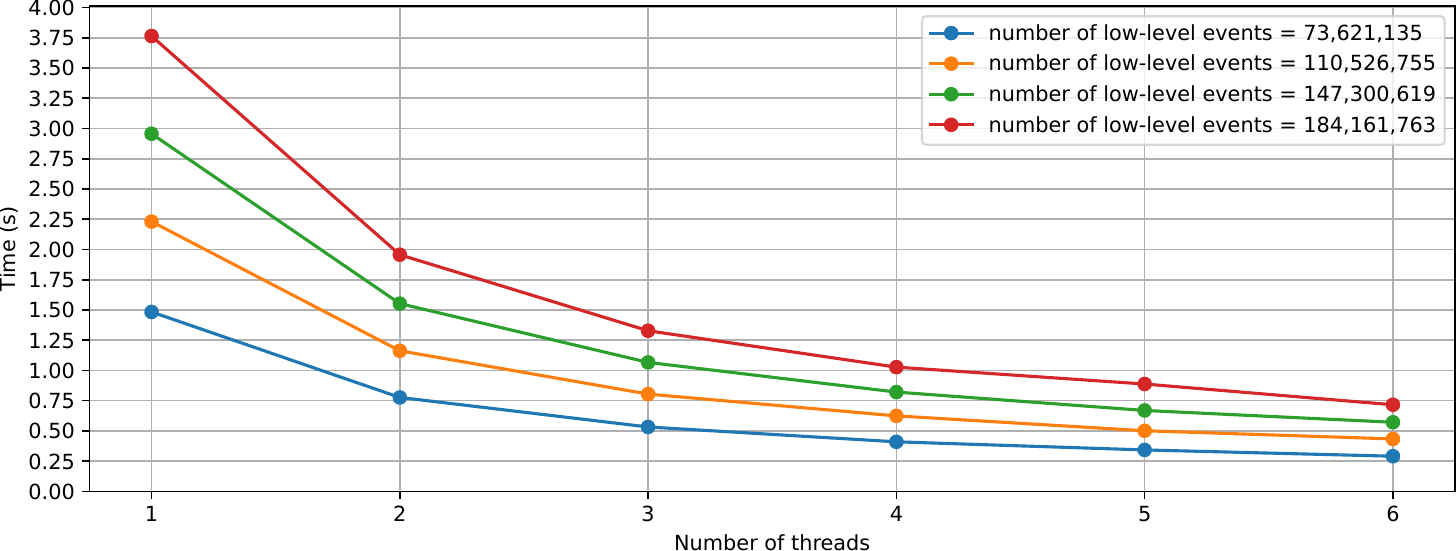}
    \caption{Total computation time per size of the datasets compared to the number of threads; non progressive processing. The results follow the $\frac{1}{x}$ law; see also \autoref{fig_resultsThreadsRatio}\label{fig_resultsThreadsTime}.}
\end{figure}
\begin{figure}[H]\centering
    \includegraphics[width=0.5\linewidth]{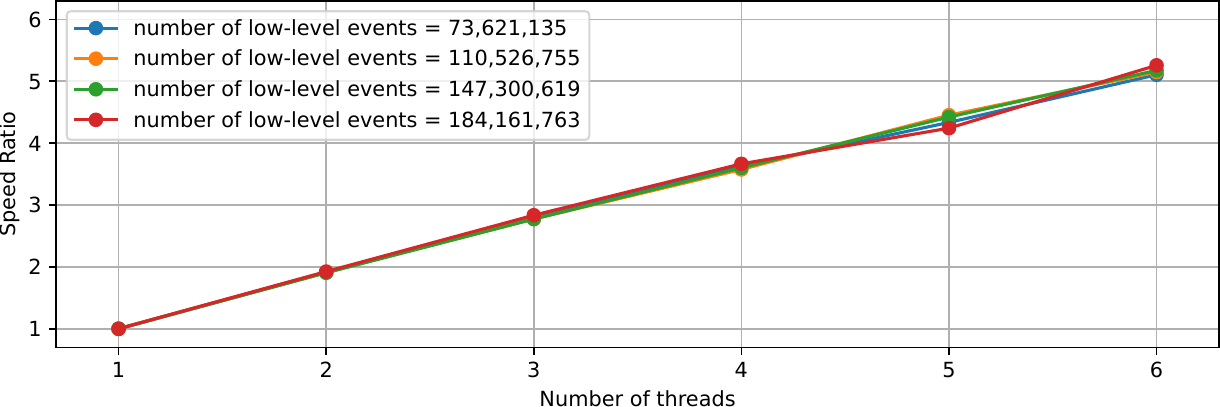}
    \caption{Computational speed per size of the dataset compared to mono-threaded computations; non-progressive environment. The speed increases linearly with the number of CPU cores.
      \label{fig_resultsThreadsRatio}}
\end{figure}

The two following figures are complementary to \autoref{fig_stability} and \autoref{fig_stabilityCINodes}.

\begin{figure}[H]\centering
    \includegraphics[width=0.7\linewidth]{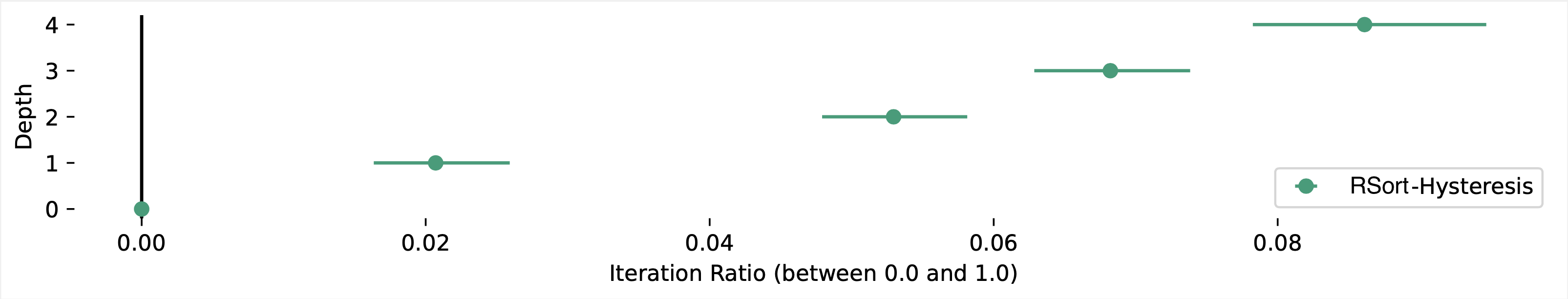}
    \caption{The 95\% CIs, computed by Bootstrapping (BCA), of how fast the \emph{Hysteresis Sort} stabilizes nodes compared to \emph{RSort} by depth.\label{fig_PWstabilityCINodes}}
\end{figure}

\begin{figure}[H]\centering
    \includegraphics[width=0.7\linewidth]{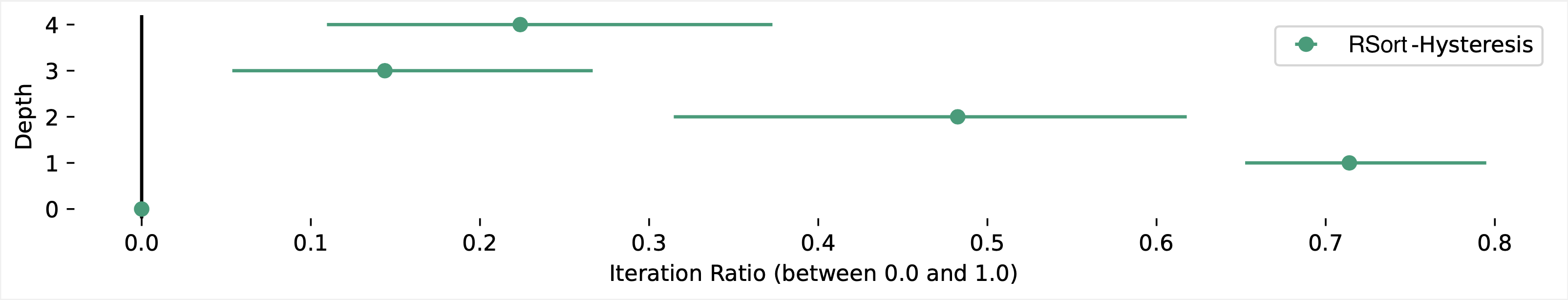}
    \caption{The 95\% CIs, computed by Bootstrapping (BCA), of how fast the \emph{Hysteresis Sort} stabilizes the tree compared to \emph{RSort} by depth.\label{fig_PWstability}}
\end{figure}

\begin{figure}[H]\centering
    \includegraphics[width=0.8\linewidth]{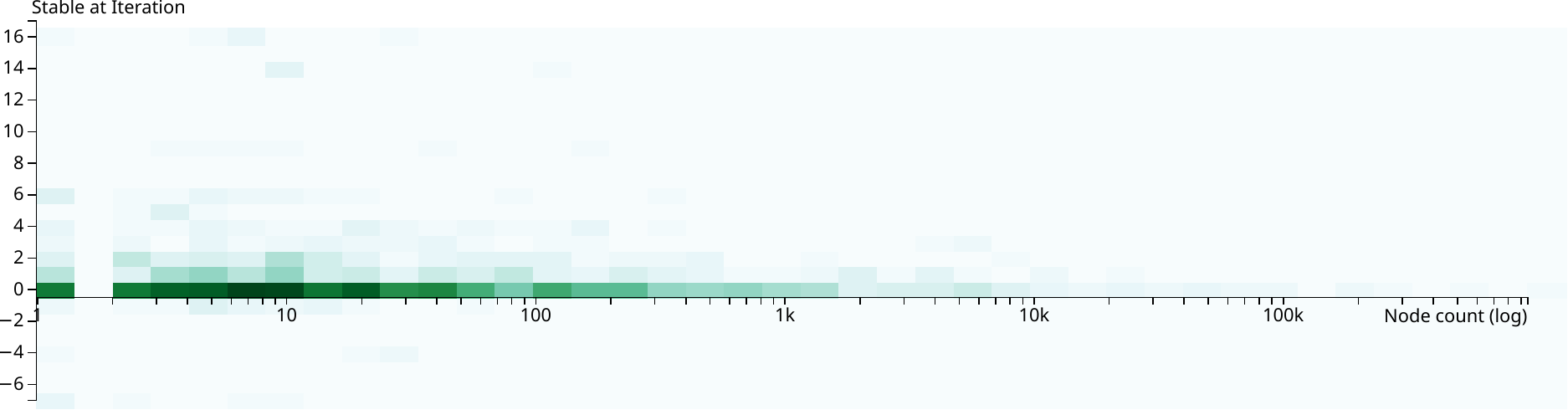}
    \caption{A heatmap showing how soon (in number of iterations) the \emph{Hysteresis Sort} stabilizes nodes (categorized per their log-scaled frequency) compared to the \emph{RSort}. Positive values along the y-axis mean that the \emph{Hysteresis Sort} stabilizes nodes sooner than the \emph{RSort}. The heatmap is computed for $\textit{threshold}=0$ and $\textit{chunkSize}=100K$, and considers only nodes whose depths are equal to or lower than 4, counting from 0.
      We use \texttt{D3.js} and its BuGn color scale.
    \label{fig_heatmap}}
\end{figure}

\begin{figure}[H]\centering
    \includegraphics[width=0.7\linewidth]{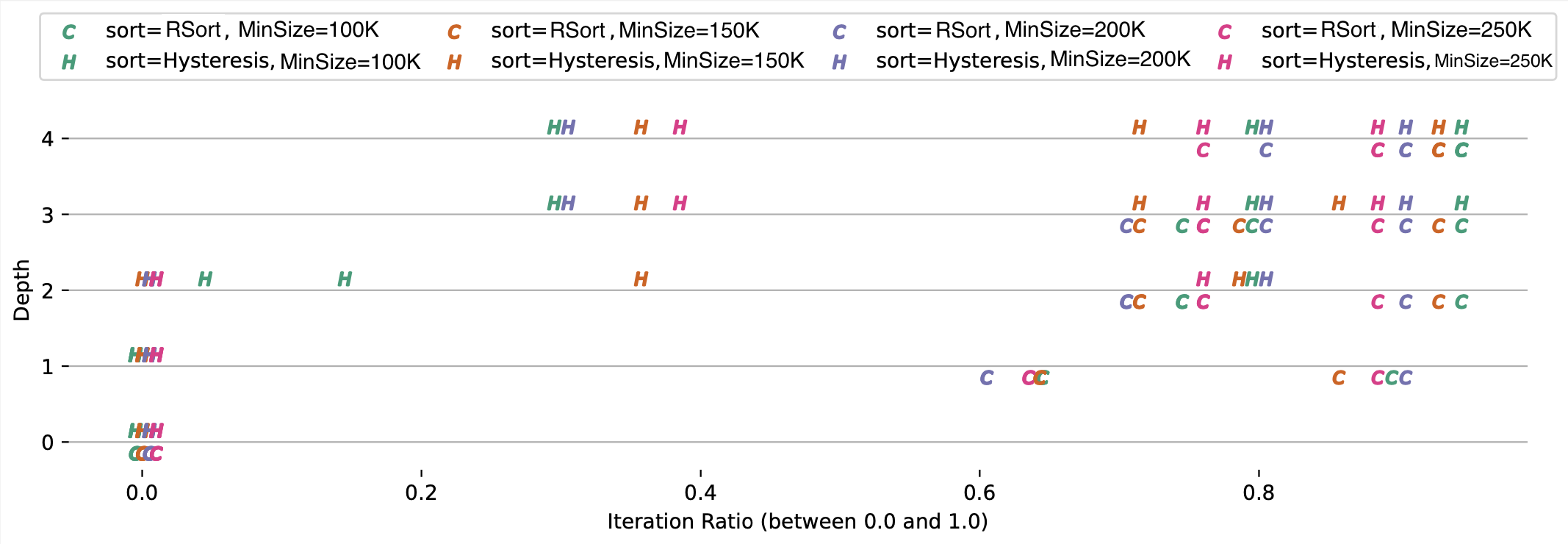}
    \caption{The stability of the tree by depth and \textit{chunkSize} values. We applied a small offset on the x-axis to distinguish overlapping points of two different \textit{chunkSize} values. Each category has $3$ \textit{MinSize} values. The figure does not highlight clusters based on \textit{chunkSize}, suggesting that this variable has no effect on how fast the tree stabilizes itself.\label{fig_perChunkSize}}
\end{figure}

\subsection*{Other Application Domains}
\label{other_domains}
\am{ParcoursVis has been designed to visualize health records, but could be used for other kinds of event sequence data. 
We adapted it to visualize the \texttt{MSNBC.com} Anonymous Web Data, the largest public dataset mentioned in the related work~\cite{DBLP_journals/tvcg/VrotsouN19}; it was retrieved from the UCI Machine Learning Repository~\cite{uci-ml}. It contains 989,925 sequences of website visits, with 18 types of web pages, i.e., event types, and the average number of visits per user is (average sequence length) 5.7.}

\begin{figure}[H]
    \centering
    \includegraphics[width=\linewidth]{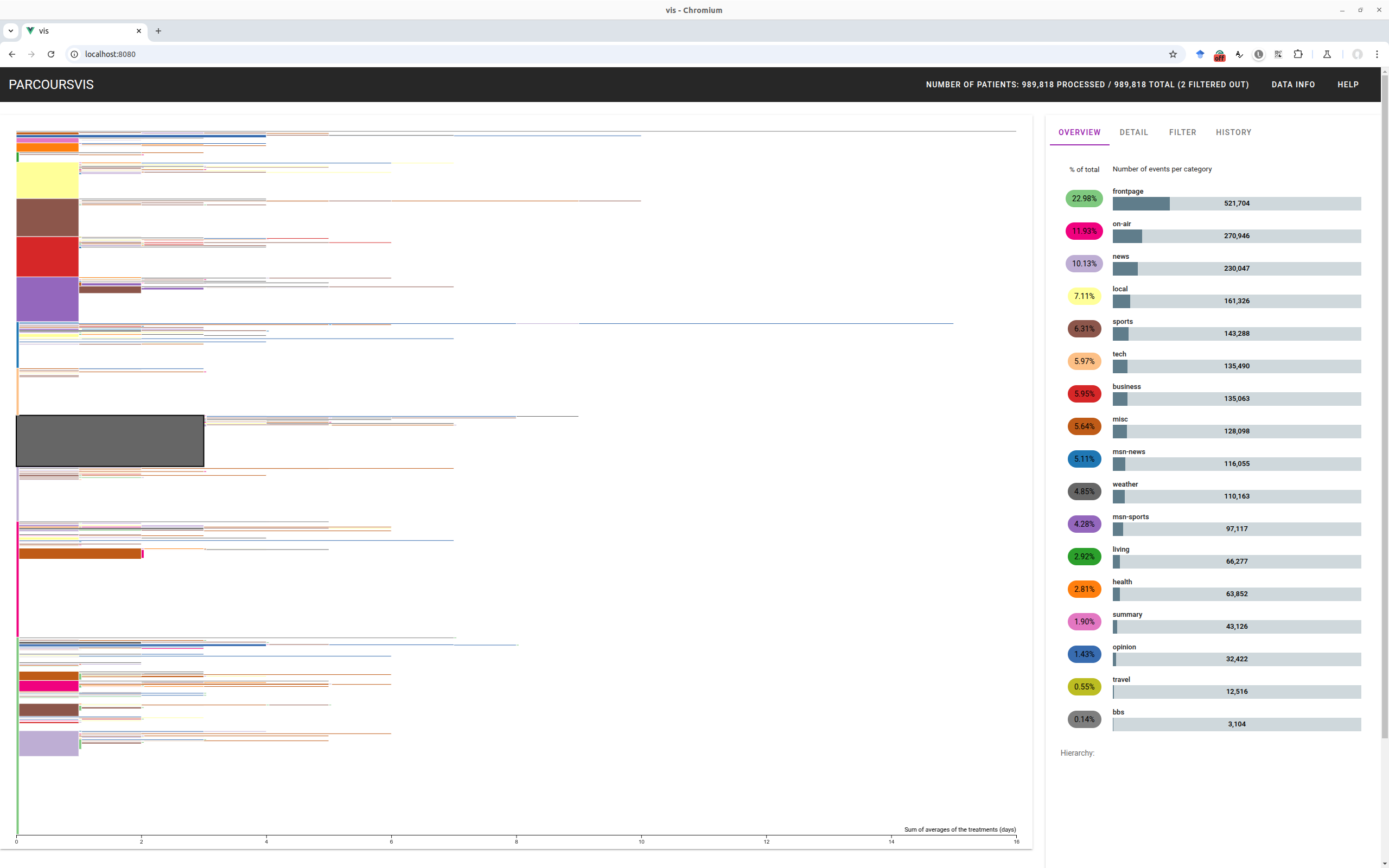}
    \caption{\am{The MSNBC.com Anonymous Web Site dataset containing about 1 million web page visits visualized with ParcoursVis.}}
    \label{fig_msnbc}
\end{figure}

\am{Converting it to ParcoursVis's format and adapting ParcoursVis took about one hour, mostly to write a script to transform the file format into ParcoursVis's CSV format and specify the colors for the events. The result is shown at \autoref{fig_msnbc} and ParcoursVis is usable without noticeable latency.
A deeper adaptation to this application domain would require changing the vocabulary shown by ParcoursVis (it still refers to the visitors as ``patients'' and to the visits as ``treatments'' in the figure), dealing with time (the dataset does not contain time information), and adapting the details and filter panels to use relevant attributes. The core visualization would remain the same.}

\end{document}